\documentclass[aip,jcp,amsmath,amssymb,reprint,superscriptaddress,longbibliography]{revtex4-1}
\usepackage[dvipdfmx]{graphicx}
\usepackage{bmpsize}
\usepackage{txfonts}
\usepackage{epsf}
\usepackage{bm}
\usepackage{lipsum}

\usepackage{color}

\begin{document}
\title{Learning reaction coordinates via cross-entropy minimization: Application to alanine dipeptide}

\author{Yusuke Mori}
\affiliation{Division of Chemical Engineering, Department of Materials Engineering Science, Graduate School of Engineering Science, Osaka University, Toyonaka, Osaka 560-8531, Japan}

\author{Kei-ichi Okazaki}
\email{keokazaki@ims.ac.jp}
\affiliation{Institute for Molecular Science, Okazaki, Aichi 444-8585, Japan}

\author{Toshifumi Mori}
\email{mori@ims.ac.jp}
\affiliation{Institute for Molecular Science, Okazaki, Aichi 444-8585, Japan}
\affiliation{The Graduate University for Advanced Studies, Okazaki, Aichi 444-8585, Japan}

\author{Kang Kim}
\email{kk@cheng.es.osaka-u.ac.jp}
\affiliation{Division of Chemical Engineering, Department of Materials Engineering Science, Graduate School of Engineering Science, Osaka University, Toyonaka, Osaka 560-8531, Japan}
\affiliation{Institute for Molecular Science, Okazaki, Aichi 444-8585, Japan}

\author{Nobuyuki Matubayasi}
\email{nobuyuki@cheng.es.osaka-u.ac.jp}
\affiliation{Division of Chemical Engineering, Department of Materials Engineering Science, Graduate School of Engineering Science, Osaka University, Toyonaka, Osaka 560-8531, Japan}

\date{\today}

\begin{abstract}
We propose a cross-entropy minimization method for finding the
 reaction coordinate from a large number of collective variables in 
 complex molecular systems.
This method is an extension of the likelihood maximization approach
 describing 
 the committor function with a sigmoid.
By design, 
the reaction coordinate as a function of various collective variables is optimized 
such that the distribution of the
 committor $p_\mathrm{B}^*$ values generated from molecular dynamics
 simulations can be described in a sigmoidal manner.
We also introduce the $L_2$-norm regularization used in the
 machine learning field to
 prevent overfitting when the number of considered collective variables 
 is large.
The current method is applied to study the isomerization of alanine
 dipeptide in vacuum, where 45 dihedral angles are used as candidate variables.
The regularization parameter is determined by cross-validation 
 using training and test datasets.
It is demonstrated that the optimal reaction coordinate involves important dihedral angles,
 which are consistent with the
 previously reported results.
Furthermore, 
the points with $p_\mathrm{B}^*\sim 0.5$ clearly indicate a separatrix
 distinguishing reactant and product states
 on the potential of mean force using the extracted dihedral angles.
\end{abstract}

\maketitle

\section{Introduction}
\label{sec:introduction}
Characterizing the free energy landscape
of complex molecular
systems is important for understanding the underlying mechanism of the dynamical
processes such as protein
isomerizations.~\cite{Chipot:2007wt,Zuckerman:2010vu}
The potential of mean force (PMF) has been utilized to describe
the complex landscape as a function of an \textit{a priori} selected small number of collective variables (CVs).
Various enhanced simulation techniques, \textit{e.g.}, 
umbrella sampling~\cite{Torrie:1977hs}, replica exchange
method~\cite{Sugita:1999cl}, and metadynamics~\cite{Laio:2002ft}, 
have been developed to obtain PMFs
efficiently.

The CV generally denotes a variable as a function of the molecular
conformation of the system.
Examples are distance and angle variables characterizing molecular structures.
Stable states, \textit{i.e.}, reactant and product, are energetically distinguished by
the saddle point of the PMF profile.
If the saddle point plays a role of 
the transition state (TS) within the framework of transition state
theory, the selected CVs serve as the reaction coordinates (RCs).~\cite{Peters:2017tf}
It is however non-trivial to find the relevant RCs from a large number of CVs.
Most importantly, 
the position of the saddle point is strongly affected by the choice of CVs.
This indicates that it is necessary to rigorously examine whether the
obtained PMF profile can predict the TS separating stable states.

The committor analysis is the statistical method to find good
RCs from the transition paths sampled by molecular dynamics (MD) simulations.~\cite{Bolhuis:2002ew}
Let A and B denote the reactant and product states that are divided by the TS, respectively.
Here, the ``committor'' $p_\mathrm{B}(\mathbf{x})$ is defined as the probability
of the trajectories that reach the state B prior to the state A starting
from a conformation $\mathbf{x}$ with the Maxwell--Boltzmann distributed
velocity (typically on the order of 100
trajectories).
If this $\mathbf{x}$ is located at the TS, 
$p_\mathrm{B}=1/2$ because of equal probability reaching A and B.
In other words, the TS can be defined as a set of conformations such that
$p_\mathrm{B} = 1/2$ using a good RC $r(\mathbf{x})$.
Practically, 
the committor distribution $p(p_\mathrm{B})$ obtained from large numbers
of initial points near the TS has a sharp peak at
$p_\mathrm{B} =1/2$.
There have been many applications of the committor
distribution test when examining the quality of the chosen
coordinate.~\cite{Du:1998iu, Geissler:1999kj, Bolhuis:2000eo,
Dellago:2002db, Hagan:2003kt, Hummer:2004ib, Pan:2004eu, Ma:2005jh, Ren:2005kb, Rhee:2005bz, E:2005ho,
Berezhkovskii:2005er, Best:2005cx, Moroni:2005es, Peters:2006cv,
Branduardi:2007ib, Quaytman:2007vo,
Antoniou:2009gg, Peters:2010ku}

\begin{figure}[t]
\centering
\includegraphics[width=0.4\textwidth]{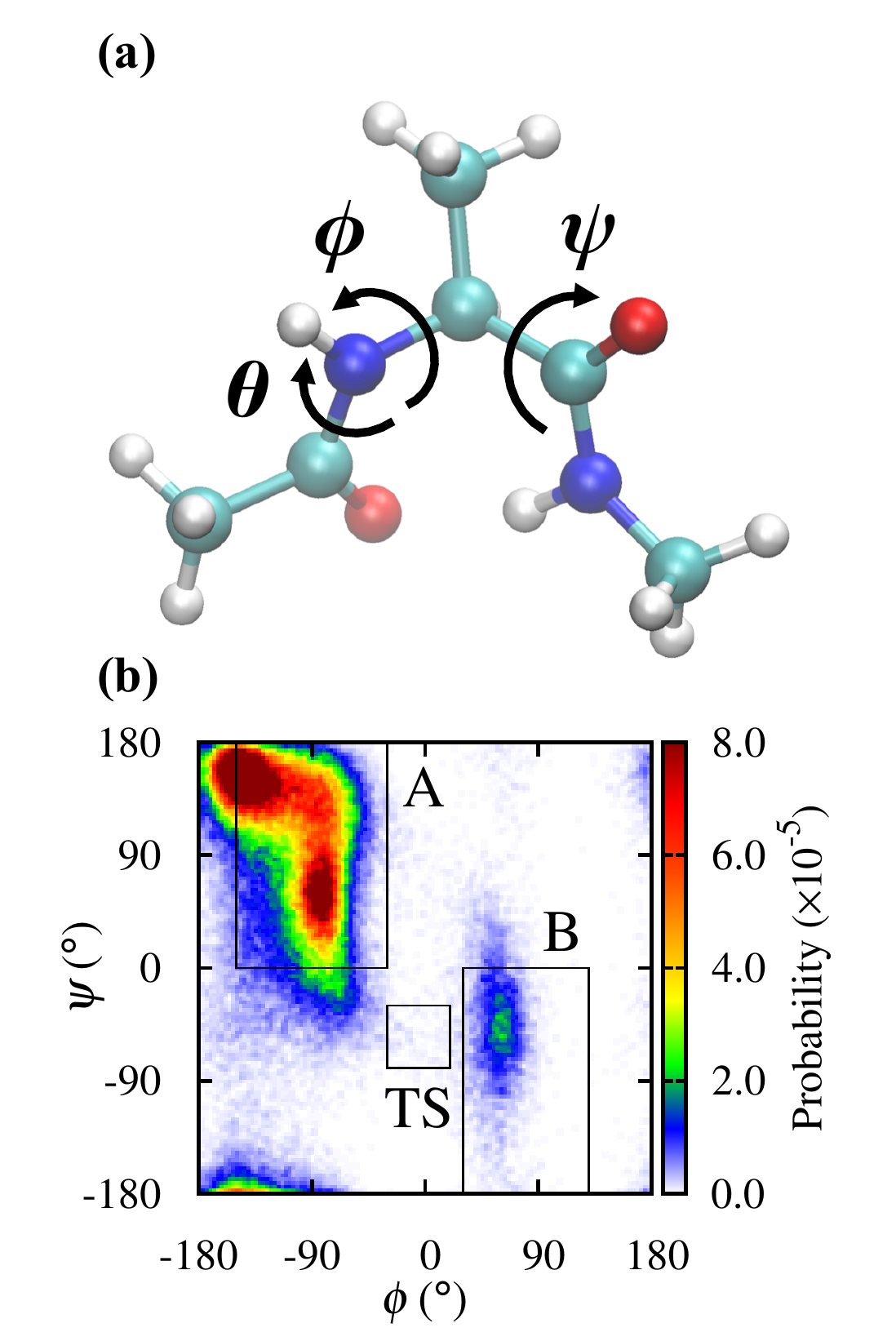}
\caption{(a) Schematic representation of the alanine dipeptide molecule
 and its major dihedral angles,
 $\phi\left(\mathrm{C-N-C_{\alpha}-C}\right)$,
 $\psi\left(\mathrm{N-C_{\alpha}-C-N}\right)$, and 
 $\theta\left(\mathrm{O-C-N-C_{\alpha}}\right)$. 
(b) Ramachandran plot of alanine dipeptide in vacuum.  
The regions
 described in boxes are defined as A: ($-150^{\circ} \le \phi \le
 -30^{\circ}$, $0^{\circ} \le \psi \le 180^{\circ}$), B: ($30^{\circ} \le
 {\phi} \le 130^{\circ}$, $-180^{\circ} \le {\psi} \le 0^{\circ}$), and TS:
 ($-30^{\circ} \le {\phi} \le 20^{\circ}$, $-80^{\circ} \le {\psi} \le
 -30^{\circ}$).
}
\label{fig1}
\end{figure}

In the seminal work by Bolhuis \textit{et al.}, 
the committor analysis has been applied to the
isomerization of alanine dipeptide.~\cite{Bolhuis:2000eo}
For characterizing protein isomerizations, the Ramachandran
plot, which is a histogram of backbone dihedral angles $\phi$ and
$\psi$ of amino acids, has conventionally been visualized (see
Fig.~\ref{fig1}(a) for the definition of $\phi$ and $\psi$).
In vacuum, two energetically stable states, the $\beta$-sheet structure
(state A) and the left-handed $\alpha$-helix structure (state B), are
characterized by this plot (see Fig.~\ref{fig1}(b) for states A and B).
However, Bolhuis \textit{et al.} reported that an additional dihedral angle $\theta$ is
required to appropriately obtain the proper committor distribution (see also 
Fig.~\ref{fig1}(a) for the definition of $\theta$).
That is, the Ramachandran plot using two angles $\phi$ and $\psi$
can distinguish the two states A and B, but is not capable of
predicting the TS properly.

The committor analysis for extracting appropriate RCs has been done via a
``trial-and-error'' approach based on physical intuition.
Remarkably, Ma and Dinner have developed the genetic neural network method, which was
applied to committor values evaluated for various conformations.~\cite{Ma:2005jh}
It was demonstrated that the optimized CVs for describing the 
committor distribution showing the peak at $p_\mathrm{B}=1/2$ involve the
dihedral angle $\theta$ in vacuum.
This results is consistent with the previous study by Bolhuis \textit{et al.}~\cite{Bolhuis:2000eo}
The importance of the angle $\theta$ has also been discussed by Ren, \textit{et al.}~\cite{Ren:2005kb}

Overall, developing reliable and efficient methods to identify 
RCs is still a demanding task in MD
simulations.~\cite{Peters:2010hm, Li:2014fp, Wales:2015dl,
Peters:2016by, Banushkina:2016hi, Sittel:2018gs, Sultan:2018fo, Jung:2019td,
Noe:2020ic, Sidky:2020im}
Peters, \textit{et. al.}, have recently developed an approach using 
the likelihood maximization method for finding RCs.~\cite{Peters:2007et}
In their method, the likelihood as a function of the committor value 
was introduced, and combined with an aimless shooting algorithm, 
which is a variation of the transition path sampling method.~\cite{Peters:2006iz}
The aimless shooting generates a binary outcome with respect to the
committor value, \textit{i.e.},
$p_\mathrm{B}^*=0$ or 1, for each trajectory from one shooting point.
The committor was modeled as the sigmoid function
$p_\mathrm{B}(r)=[1+\tanh(r)]/2$, and 
the likelihood maximized using those outcomes led to 
the RC $r$ by optimizing linear combinations of
the CVs of sampled shooting points.~\cite{Peters:2007et}
The likelihood maximization method has widely been utilized for finding
the good RC in various systems.~\cite{Beckham:2007jz,
Beckham:2008hh, Vreede:2010ig, Lechner:2010du, Pan:2010ky, Beckham:2011dw,
Lechner:2011cv, Peters:2012cv, Xi:2013en, Jungblut:2013hn, Mullen:2014cl,
Mullen:2015jb, Lupi:2016dda, Jung:2017cd,
Joswiak:2018jl, DiazLeines:2018dh, Okazaki:2019jx}

In this study, 
we propose a refined approach for identifying the RC using dataset of 
the pre-evaluated committor value $p_\mathrm{B}^*$ that varies continuously
from 0 to 1.
This method requires more \textit{a priori} calculations for $p_\mathrm{B}^*$ 
than the binary outcomes.
However, 
the continuous nature of the committor will provide a more accurate statistics for the RC.
We illustrate that the likelihood maximization is naturally extended to the
cross-entropy minimization.
Note that these approaches, corresponding to the Logistic regressions in the 
machine learning literature, often suffer from overfitting.~\cite{Bishop:2006ui}
To prevent overfitting, we introduce the $L_2$-norm regularization to 
the cross-entropy minimization.

The presented cross-entropy minimization method is applied to study the
isomerization of alanine dipeptide in vacuum.
We use all dihedral angles of the molecule as candidate CVs and
perform the cross-entropy minimization with the committor values
$p_\mathrm{B}^*$ to search the best RC representing the TS.
The regularization parameter is heuristically determined by 
cross-validation using training and test datasets.
Finally, we examine the validity of the optimized coordinate by plotting
the committor distributions as a function of characteristic CVs.

The remaining sections of this paper are organized as follows. 
Sec. II
describes the formalism of the cross-entropy minimization as an
generalization of the likelihood maximization.
We also introduce the $L_2$-norm regularization into the objective function.
In Sec. III, we present the computational details with regard to the
generation of the $p_\mathrm{B}^*$ data and cross-entropy minimization.
In Sec. IV, the numerical results and discussions are described.
Finally, our conclusions are drawn in Sec. V.

\section{Theory}

\subsection{Likelihood maximization and cross-entropy minimization}

We start from $N$ snapshots of the system that are sampled from the 
path connecting the reactant A and product B.
We describe each snapshot $k$ by $M$ CVs $q_i(\mathbf{x}_k)$, which are functions of the
Cartesian coordinates $\mathbf{x}_k$.
The committor calculated at each point from multiple short
simulations is denoted as $p_\mathrm{B}^*\left(\mathbf{x}_k\right)$.

We aim at obtaining a RC
that can describe the change of committor distribution
$p_\mathrm{B}^*$ in a sigmoidal manner.
To this end, we define the CV vector
$\mathbf{q}(\mathbf{x}_k)=(1, q_{1}(\mathbf{x}_k), \cdots,
q_{M}(\mathbf{x}_k))$ 
and corresponding coefficients $\bm{\alpha}=(\alpha_{0}, \alpha_{1},\cdots, \alpha_{M})$.
Note that $\mathbf{q}$ is ($M$+1)-dimensional due to the bias term ($q_0=1$).
We describe the trial function $r\left(\mathbf{q}\left(\mathbf{x}_k\right)\right)$ as a linear combination of the CVs as
\begin{align}
r\left(\mathbf{q}\left(\mathbf{x}_k\right)\right)
= \bm{\alpha}\cdot\mathbf{q}\left(\mathbf{x}_k\right)
= \sum^{M}_{m=1}\alpha_{m}q_{m}\left(\mathbf{x}_k\right) + \alpha_0.
\label{eq:linear}
\end{align}
We assume that, in the ideal case, the committor $p_\mathrm{B}$ changes from 0 to 1 following the sigmoid function defined by
\begin{align}
p_\mathrm{B}\left(r\left(\mathbf{q}\left(\mathbf{x}_k\right)\right)\right)
= \frac{1+\tanh\left(r\left(\mathbf{q}\left(\mathbf{x}_k\right)\right)\right)}{2}.
\label{eq:sigmoid}
\end{align}
Using Eq. (\ref{eq:sigmoid}), the Likelihood function $\mathcal{L}\left(\bm{\alpha}\right)$ can be defined as
\begin{align}
\mathcal{L}\left(\bm{\alpha}\right)
= \prod_{\mathbf{x}_k\rightarrow \mathrm{B}}p_\mathrm{B}\left(r\left(\mathbf{q}\left(\mathbf{x}_k\right)\right)\right) 
 \times \prod_{\mathbf{x}_k\rightarrow \mathrm{A}}\left(1-p_\mathrm{B}\left(r\left(\mathbf{q}\left(\mathbf{x}_k\right)\right)\right)\right),
\label{eq:likelihood}
\end{align}
which was originally introduced by Peters \textit{et al.}~\cite{Peters:2007et}
Here, $\mathbf{x}_k\rightarrow \mathrm{B}$ and $\mathbf{x}_k\rightarrow \mathrm{A}$
indicate the trajectories starting from point $\mathbf{x}_k$ that ends in state B and A, respectively.
By taking the logarithmic form of Eq. (\ref{eq:likelihood}), we obtain
\begin{align}
\ln{\mathcal{L}}\left(\bm{\alpha}\right)
= \sum_{\mathbf{x}_k\rightarrow \mathrm{B}}\ln{p_\mathrm{B}\left(r\left(\mathbf{q}\left(\mathbf{x}_k\right)\right)\right)} 
+ \sum_{\mathbf{x}_k\rightarrow \mathrm{A}}\ln{\left[1-p_\mathrm{B}\left(r\left(\mathbf{q}\left(\mathbf{x}_k\right)\right)\right)\right]}.
\label{eq:log_likelihood}
\end{align}
While each point $\mathbf{x}_k$ has a fractional probability to reach
either state A or B, Eq. (\ref{eq:log_likelihood}) can only account for
each point in a binary manner to state A
($p_\mathrm{B}^*\left(\mathbf{x}_k\right)=0$) or B
($p_\mathrm{B}^*\left(\mathbf{x}_k\right)=1$).
To make use of the continuous nature of the committor 
obtained directly, we extend Eq. (\ref{eq:log_likelihood}) to
\begin{align}
&\mathcal{H}\left(p_\mathrm{B}^*, p_\mathrm{B}\right)\nonumber\\
&\quad = -\sum_{k=1}^{N} p_\mathrm{B}^*\left(\mathbf{x}_k\right)\ln{p_\mathrm{B}\left(r\left(\mathbf{q}\left(\mathbf{x}_k\right)\right)\right)} \nonumber \\
&\quad\quad - \sum_{k=1}^{N}
 \left(1-p_\mathrm{B}^*\left(\mathbf{x}_k\right)\right)\ln{\left[1-p_\mathrm{B}\left(r\left(\mathbf{q}\left(\mathbf{x}_k\right)\right)\right)\right]}, 
\label{eq:cross_entropy}
\end{align}
which is equivalent to the cross-entropy.
Note that Eq. (\ref{eq:cross_entropy}) is derived from the
Kullback--Leibler divergence in Ref.~\onlinecite{Mori:2020fs}.
Equations (\ref{eq:log_likelihood}) and (\ref{eq:cross_entropy}) are
equivalent with the opposite sign when $p_\mathrm{B}^*$ is
binary:
\begin{align}
p_\mathrm{B}^*
= \left\{\begin{array}{ll}
  0 & \left(\mathbf{x}_k \rightarrow \mathrm{A}\right) \\
  1 & \left(\mathbf{x}_k \rightarrow \mathrm{B}\right)
  \end{array}\right.
\label{eq:bimodal}
\end{align}
Thus, the likelihood maximization is generalized to the cross-entropy
minimization, considering the continuous nature of the committor.
Note that $\mathcal{H}(p_\mathrm{B}^*, p_\mathrm{B}) \ge
\mathcal{H}(p_\mathrm{B}^*)\equiv\mathcal{H}(p_\mathrm{B}^*,
p_\mathrm{B}=p_\mathrm{B}^*)$, 
where $\mathcal{H}(p_\mathrm{B}^*)$ sets the lower bound of the cross-entropy.

\subsection{$\bm{L_2}$-norm regularization}

When the number of CVs used to describe the trial function
$r\left(\mathbf{q}\left(\mathbf{x}_k\right)\right)$ is large, resulting
reaction coordinate via the cross-entropy minimization can overfit the input data. 
To avoid overfitting, we introduced a technique called
regularization that considers a penalty term in the objective
function. In particular, we used the $L_2$-norm regularization.~\cite{Bishop:2006ui}
The objective function with the regularization is,
\begin{align}
\mathcal{H}\left(\bm{\alpha}\right) = \mathcal{H}\left(p_\mathrm{B}^*, p_\mathrm{B}\right) 
+ \frac{\lambda}{2}\sum^{M}_{m=1} \alpha_{m}^{2},
\label{eq:regularization}
\end{align}
where $\lambda$ is the regularization parameter that controls the relative
weight of the penalty term.
Note that the bias term $\alpha_{0}$ is not included in the regularization.

\section{Computational details}
\subsection{Sampling global conformational space}
\label{sec:Ramachandran}

The isomerization of alanine dipeptide in vacuum was studied. 
One molecule of alanine dipeptide was placed in the 3.16 nm cubic box with the periodic
boundary conditions. 
Time step of 1 fs, neighbor-list distance of 1.5 nm, van der Waals cut-off
distance of 1.2 nm, switch function cut-off distance of 1.0 nm were
used. 
For electrostatic interaction, the particle-mesh Ewald
method was used with real-space cut-off distance of 1.2 nm. 
All covalent bonds were constrained by the LINCS algorithm.
The AMBER99SB force field was used.~\cite{Hornak:2006gx}
All simulations were conducted with GROMACS2018.1.~\cite{Abraham:2015gj}

The Ramachandran plot was generated from the replica-exchange MD (REMD)
simulation.~\cite{Sugita:1999cl}
In the setup of MD simulations, 1 ns equilibration was followed by 10 ns production
run with \textit{NVT} condition at 300 K by using the Langevin thermostat. 
In the REMD simulations, 10 replicas were
prepared in the range of 300 - 1209 K with 101 K interval. 
The exchange frequency was set to 200 fs, and the average exchange rate
was 0.3.

\subsection{Sampling conformations in transition state region}
\label{sec:aimlessshooting}

As mentioned in Sec.~\ref{sec:introduction}, Peters \textit{et al.}, proposed a variant of transition path sampling
called ``aimless shooting.''~\cite{Peters:2006iz}
In this method, trajectories are generated with freshly sampled momenta
from the Maxwell--Boltzmann distribution from every conformation.



In this study, we conducted the two-point version of the aimless
shooting following the protocol in Ref.~\onlinecite{Peters:2007et}.
We initiated the aimless shooting from a conformation randomly chosen
from the TS region (see below and Fig.~\ref{fig1}(b) for the definition of the state). 
$\tau = 2.01$ ps and $\delta{t} =10$ fs were used.
Originally, the aimless shooting was introduced to sample
conformations near $p_\mathrm{B}^*=1/2$. 
However, as mentioned in Sec.~\ref{sec:introduction}, our purpose is to sample points that 
uniformly cover committor $p_\mathrm{B}^*$ values from 0 to 1.
For this, we incorporated the shooing point even if the trajectory was rejected.
We sampled 2,000 shooting points in total (accepted and rejected trajectories), which are divided equally into
training and test datasets.
From each point, we
quantified $p_\mathrm{B}^*$ by running 1 ps MD simulations 100 times with
random velocities from the Maxwell-Boltzmann distribution at 300 K. 

\begin{figure}[t]
\centering
\includegraphics[width=0.4\textwidth]{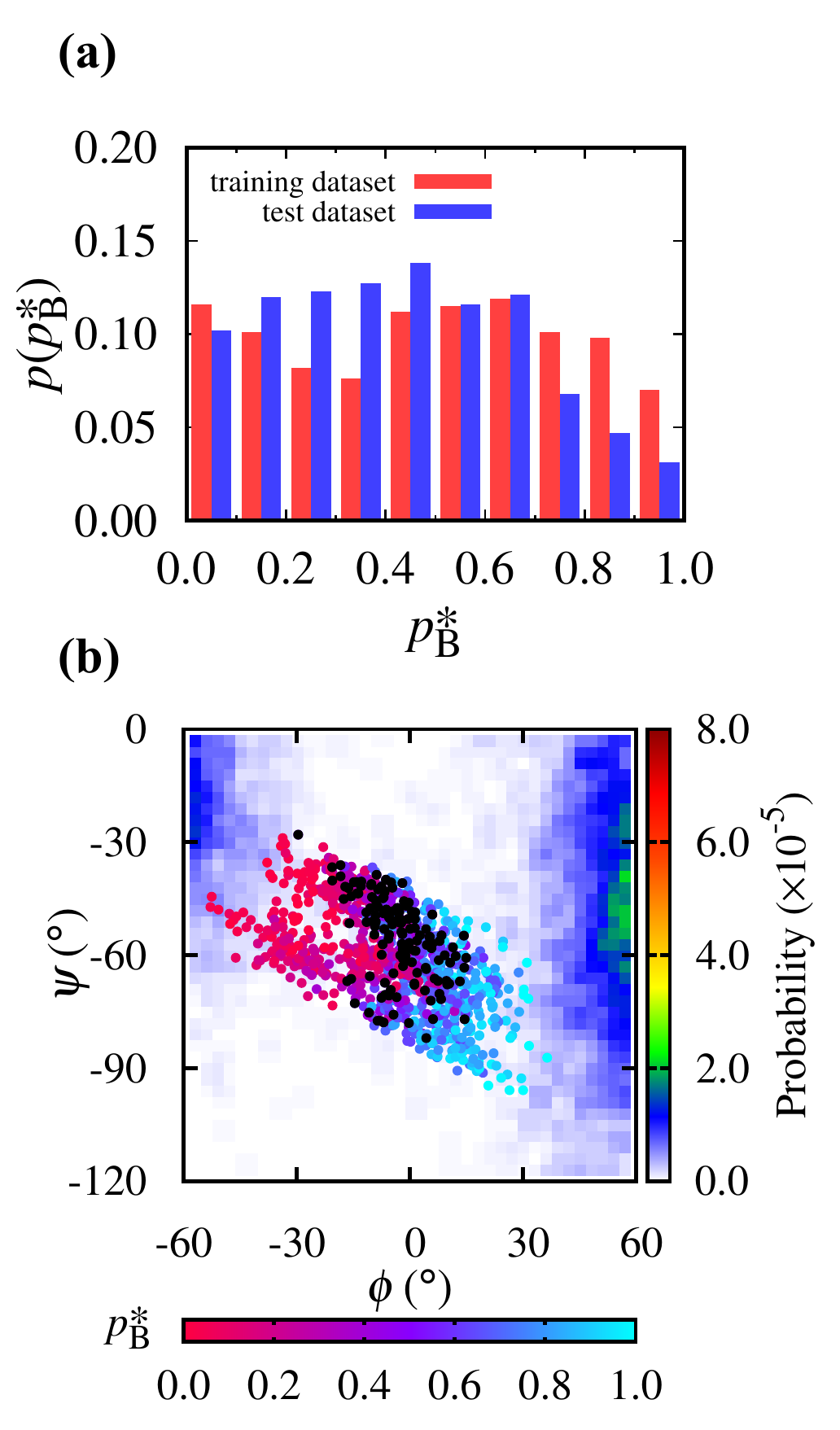}
\caption{(a) Probability of committor value $p_\mathrm{B}^*$ for the training (red) and test (blue)
 datasets. Each dataset consists of 1,000 points, and $p_\mathrm{B}^*$ for each
 point is calculated from 100 trajectories.
(b) Distribution of the training data points plotted on the Ramachandran
 plot of Fig.~\ref{fig1}(b).
The points are colored by the $p_\mathrm{B}^*$ values given in the bottom color bar.
In addition, the points with $p_\mathrm{B}^* \sim 0.5$ ($0.45\le p_\mathrm{B}^* \le 0.55$) are marked in black dots.
}
\label{fig:TPS_distribution}
\end{figure}

\subsection{Reaction coordinate optimization via cross-entropy minimization}
\label{sec:crossentropy}

Using the $p_\mathrm{B}^*$ values, we performed the cross-entropy minimization.
We considered 45 dihedral angles (see Fig.~S1 and Table S1 of Supplementary Material).
These dihedral angles were transformed into cosine and sine forms, considering the periodicity. 
Thus, the dimension of $\bm{\alpha}$ is 91 ($M=90$ plus 1 bias term).
The steepest descent method was used to update the coefficients
$\bm{\alpha}$ as,
\begin{eqnarray}
\bm{\alpha}^{(n+1)} = \bm{\alpha}^{(n)} - \gamma\nabla\mathcal{H}\left(\bm{\alpha}^{(n)}\right),
\label{eq:gradient}
\end{eqnarray}
where $\bm{\alpha}^{(n)}$ and $\bm{\alpha}^{(n+1)}$ are the parameters at
the $n$-th and ($n$+1)-th steps, respectively.
$\nabla\mathcal{H}(\bm{\alpha}^{(n)})$ represents the gradient at
the $n$-th step and $\gamma$ is the step size which was fixed to $10^{-5}$.
The optimal $\bm{\alpha}$ was determined when the norm of
$\nabla\mathcal{H}\left(\bm{\alpha}\right)$ becomes less than
$\varepsilon = 10^{-3}$.
The regularization parameter was chosen as $\lambda =0$, 0.1, 0.5, 1, 10, and 100.
To check the robustness of the optimization, we ran 10 optimization trials from the
initial coefficients $\alpha_i$ that are randomly sampled from the 
range of $-0.1 \le \alpha_i \le 0.1$.


\begin{figure}[t]
\centering
\includegraphics[width=0.4\textwidth]{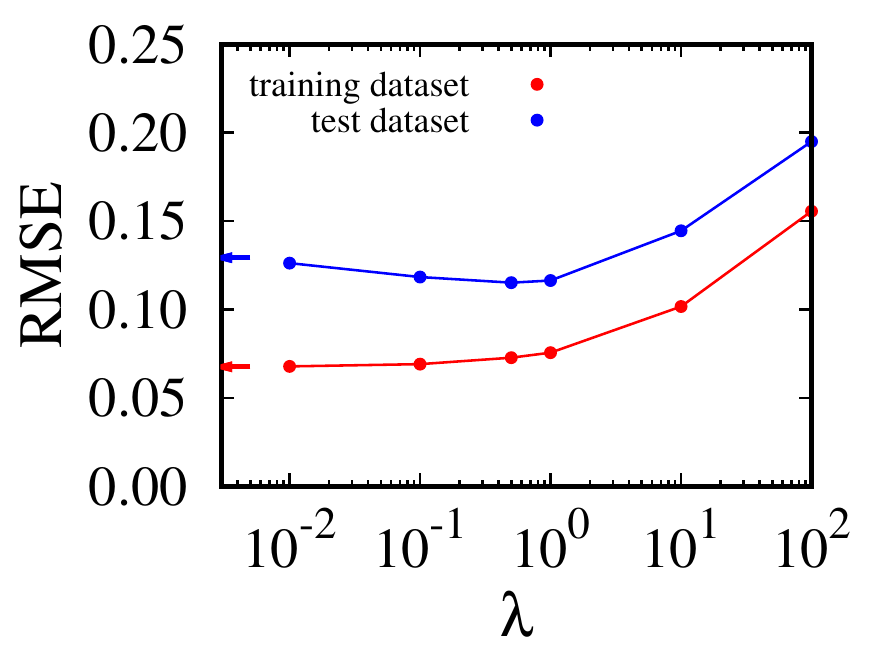}
\caption{RMSEs of the training (red) and test (blue) datasets as a function of the
 regularization parameter $\lambda$.
RMSE values of $\lambda=0$ are indicated by the arrows.
}
\label{fig:validation}
\end{figure}

\section{Results and discussion}
\begin{figure*}[t]
\centering
\includegraphics[width=0.7\textwidth]{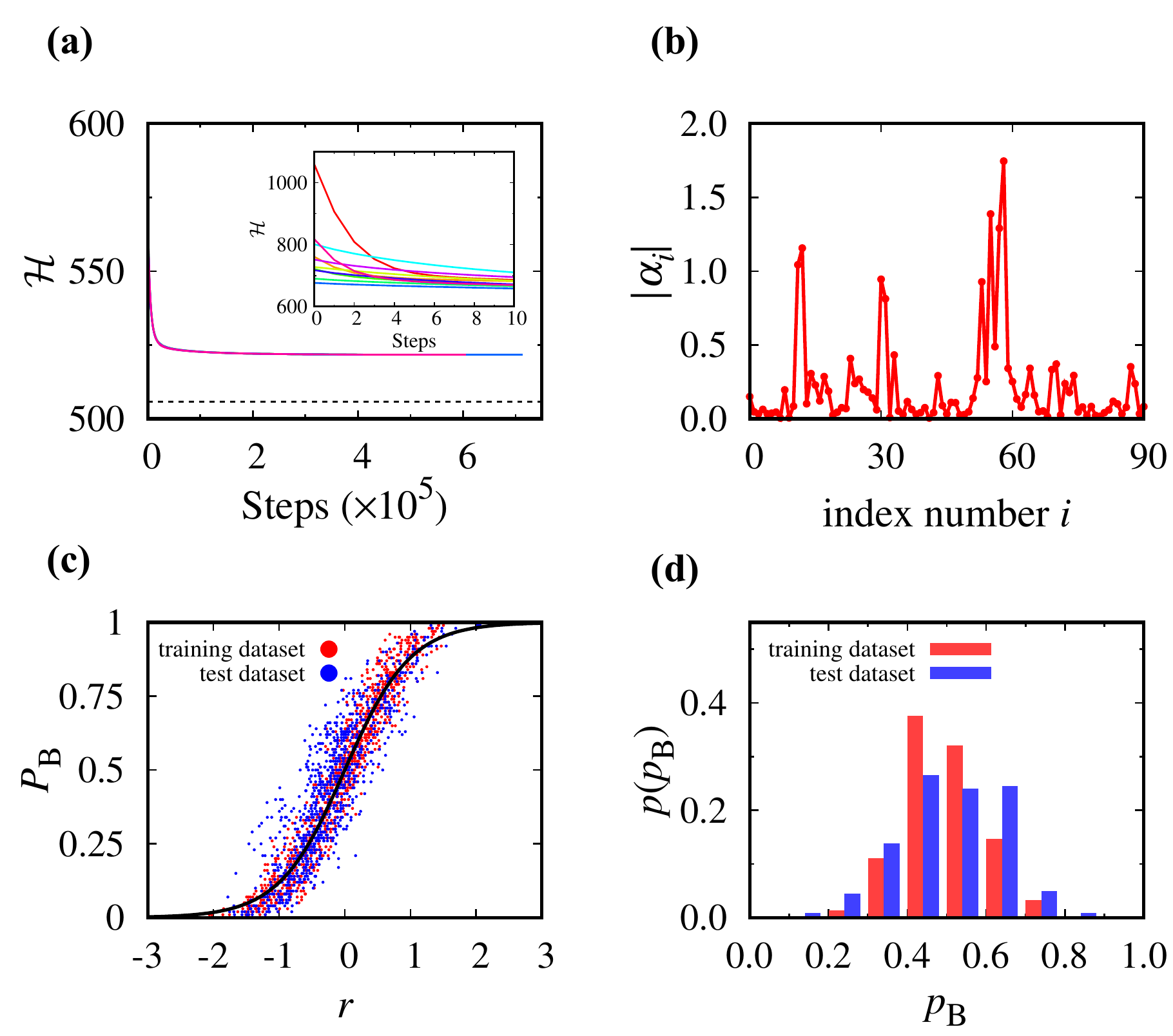}
\caption{Summary of the parameter optimization for $\lambda = 0.5$. (a)
 Changes of the cross-entropy function ($\mathcal{H}$) during the
 optimization steps (solid lines) and the ideal value
 $\mathcal{H}(p_\mathrm{B}^*)$ (black dashed line). 
The results for the 10 trials
 using different initial $\bm{\alpha}$-guesses are shown in different
 colors. 
The inset focuses on the first 10 steps, showing that
 $\mathcal{H}$ differs remarkably in the beginning but quickly converges
 to a similar value within 10 steps.
(b) Optimized coefficients ($\alpha_i$) in absolute value. 
Note that the
 coefficients are determined as an average over the 10 trials.
(c) Committor distributions of the training (red) and test (blue)
 datasets as a function of the optimized coordinate $r$. 
The sigmoid function (Eq. (\ref{eq:sigmoid})) is shown in black line.
(d) Probability of $p_\mathrm{B}$ at about the TS of $r$ ($
 -0.2 \le r \le 0.2$), where the points are extracted from the data
 shown in (c).
}
\label{fig:grad}
\end{figure*}

\begin{table}[t]
\caption{First ten dominant coefficients after optimization using
 $\lambda = 0.5$. The results are given as a mean and standard deviation
 of 10 trials starting from different initial conditions. The index
 follows the list given in Table S1 of Supplementary Material.}
\centering
\begin{ruledtabular}
      \begin{tabular}{crc}
      index & \multicolumn{1}{c}{$\alpha_i$} & \multicolumn{1}{c}{standard deviation} \\ \hline  
      $58$ & $  1.7453$  & $2.2132 \times 10^{-3}$ \\
      $55$ & $  1.3872$  & $1.4342 \times 10^{-3}$ \\
      $57$ & $ -1.2905$  & $1.3520 \times 10^{-3}$ \\
      $12$ & $  1.1562$  & $1.2566 \times 10^{-3}$ \\
      $11$ & $ -1.0431$  & $1.4347 \times 10^{-3}$ \\
      $30$ & $ -0.9451$  & $1.2216 \times 10^{-3}$ \\
      $53$ & $ -0.9275$  & $1.2669 \times 10^{-3}$ \\
      $31$ & $  0.8127$  & $1.2908 \times 10^{-3}$ \\
      $56$ & $ -0.4889$  & $1.1470 \times 10^{-3}$ \\
      $33$ & $ -0.4320$  & $1.4202 \times 10^{-3}$ \\ 
      \end{tabular}
\end{ruledtabular}
\label{table:coeff}
\end{table}

\begin{figure}[t]
\centering
\includegraphics[width=0.4\textwidth]{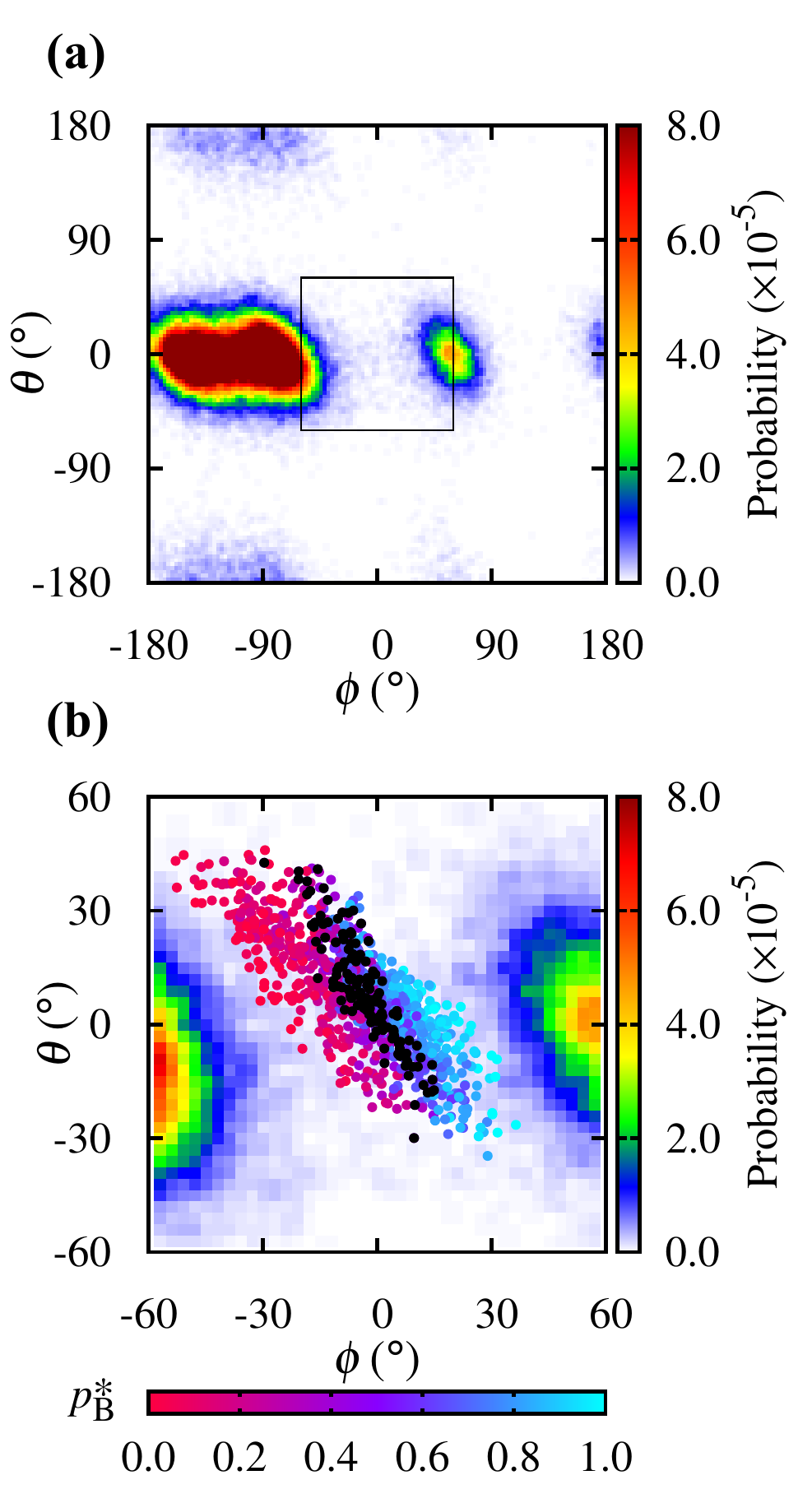}
\caption{(a) Contour plot of the probability distribution as a function of $\phi$ and $\theta$.
The probability distribution, calculated from the REMD trajectory, is
 described by a color bar on the right side of the plot.
(b) Distribution of the training data points plotted on the probability
 distribution given in the squared region of (a).
The points are colored by the $p_\mathrm{B}^*$ values given in the bottom color bar.
In addition, the points with $p_\mathrm{B}^* \sim 0.5$ ($0.45\le p_\mathrm{B}^* \le 0.55$) are marked in black dots.
}
\label{fig:shooting_point}
\end{figure}

\subsection{Training and test datasets of committor values $\bm{p_\mathrm{B}^*}$}

The Ramachandran plot obtained from the REMD trajectory is shown in Fig.~\ref{fig1}(b).
The two stable states, namely C7${}_{\mathrm{eq}}$ and
C7${}_{\mathrm{ax}}$, are found at $\phi \sim -90^{\circ}$ and $\phi \sim 60^{\circ}$, respectively.
For simplicity, hereafter we denote the C7${}_{\mathrm{eq}}$ and
C7${}_{\mathrm{ax}}$ states as A and B, respectively.
Here we
examine paths connecting states A and B, which possibly passes through
TS region at $\psi \sim -50^{\circ}$ and $\phi \sim 0^{\circ}$.
Note that these paths have also been of focus in the previous
studies.~\cite{Bolhuis:2000eo, Ma:2005jh, Ren:2005kb}
The snapshots along this path are sampled using the aimless shooting
protocol as described in Sec.~\ref{sec:aimlessshooting}.
To optimize and validate the RC, we prepared two datasets,
\textit{i.e.}, training and test, each consisting of 1,000
points.
The committor value $p_\mathrm{B}^*$ for each point was calculated by running 100 short
trajectories (see also Sec.~\ref{sec:aimlessshooting}).
Figure~\ref{fig:TPS_distribution}(a) shows the
committor distribution for the training and test datasets.
We see that the two datasets both fully cover $0 \le p_\mathrm{B}^* \le 1$ with
roughly similar probabilities.
When the points are plotted on the Ramachandran plot (shown in
Fig.~\ref{fig:TPS_distribution}(b)), we find that $\phi$ and $\psi$ can
roughly separate points reaching state A ($p_\mathrm{B} < 1/2$) and B ($p_\mathrm{B} >
1/2$).
Yet, the points with $p_\mathrm{B}^* \sim 0.5$ are spread out in the ($\phi$,
$\psi$) space without a clear ``separatrix'' ($p_\mathrm{B} = 1/2$ surface),
indicating that the two coordinates are not sufficient in characterizing
the TS.
This unclear separatrix is in accord with a rather uniform
distribution of the commmittor value $p_\mathrm{B}^*$ for the
conformations of the TS on the $\phi$-$\psi$ plane that was demonstrated via the committor analysis
in Ref.~\onlinecite{Bolhuis:2000eo}.

\subsection{Minimizing cross-entropy and determining regularization parameter}
We optimized the coefficients $\bm{\alpha}$ that minimize the
cross-entropy function $\mathcal{H}(\bm{\alpha})$
(Eq.~(\ref{eq:regularization})) using the training dataset.
To see the effect of the $L_2$-norm regularization, we changed the
regularization parameter $\lambda$ in the range of 0 to 100, and performed
the parameter optimization and validation.
The performance against the training and test datasets were measured by
the root-mean-squared-error (RMSE) between the expected
(Eq.~(\ref{eq:sigmoid})) and raw committor values, defined as
\begin{eqnarray}
\mathrm{RMSE}(\lambda) = \sqrt{\dfrac{1}{N}\sum_{k=1}^{N} \left[
							   p_B^{\ast}(\mathbf{x}_k)
							   -
							   p_B(r(\mathbf{q}(\mathbf{x}_k)))
							  \right]^2},
\end{eqnarray}
with $N = 1000$ points. 
The results of RMSEs for different choices of $\lambda$ are summarized
in Fig.~\ref{fig:validation}.
The figure shows that as $\lambda$ is increased, the RMSE of the
training data gradually increase; on the contrary, the RMSE of the test
data decreases until $\lambda \sim 1$, and starts to increase
thereafter.
Considering the balance between the performances of the training and
test datasets, the optimal choice of $\lambda$ in the current case was 
determined to be $\lambda = 0.5$.
Below, we focus on the results obtained by fixing $\lambda$ to 0.5.

\subsection{Validation of the optimized parameter set}
\label{sec:validation}

We examined the robustness of the optimization procedure using
$\lambda = 0.5$.
Figure~\ref{fig:grad}(a) shows that the cross-entropy function
($\mathcal{H}$) consistently converges to the same minimum when the
initial guess for $\bm{\alpha}$ is varied.
Figure~\ref{fig:grad}(b) gives the optimized parameters (in absolute
number), which is given as a mean of the 10 optimization trials.
The result shows that several characteristic coordinates dominate the
trial function $r(\mathbf{q}(\mathbf{x}_k))$; the raw coefficients of the
major components are summarized in Table~\ref{table:coeff}, and its full
list is shown in Table S2 of Supplementary Material.
For comparison, the results using $\lambda = 0$ and $\lambda
= 10$ are also shown in Table S3 and Table S4 of Supplementary Material,
respectively.


Using the optimized coefficients, the performance of the predictability
is tested using the test dataset.
Figure~\ref{fig:grad}(c) compares the distributions of the $p_\mathrm{B}$-value
as a function of the optimized coordinate $r$.
We see that overall the training and test datasets follow the sigmoid
function (described as a black line in Fig.~\ref{fig:grad}(c)), indicating that the optimized coordinate
does serve as a good RC for the two datasets.
We note that the test dataset tends to deviate slightly towards $p_\mathrm{B}$
value larger than the sigmoid function.
Indeed, this trend can be confirmed by looking at the probability of
$p_\mathrm{B}$ at about the TS of $r$ ($-0.2 \le r \le 0.2$), which
is given in Fig.~\ref{fig:grad}(d).
The probability show that while the distribution of $p_B$ is sharply
peaked at about $p_\mathrm{B} \sim 0.5$ for the training dataset, the peak for
the test dataset becomes broad and the center is shifted slightly
towards $p_\mathrm{B} \sim 0.6$.
Despite these small differences, the two probabilities can be
characterized by a single peak centered at $p_\mathrm{B} \sim 0.5$ and with no
points at $p_\mathrm{B} < 0.1$ and $p_\mathrm{B} > 0.9$.
The current results thus confirm that the optimal RC determined
using the training dataset is able to characterize the TS of 
the training dataset.
Note that the results corresponding to Fig.~\ref{fig:grad}(c) and
Fig.~\ref{fig:grad}(d) for $\lambda = 0$ and
$\lambda= 10$ are shown in Fig. S2 and Fig. S3 of Supplementary
Material, respectively.

\subsection{Character of the optimized reaction coordinate}
\label{sec:character}

As described in Fig.~\ref{fig:grad}(b) and Table \ref{table:coeff}, the optimal coordinate can be
characterized with a few dominant CVs.
The first two components, $\alpha_{58}$, and $\alpha_{55}$, corresponds
to the coefficient of $\sin{\phi}$ (5-7-9-15) and $\sin{\theta}$
(6-5-7-9), respectively (see also  Fig.~S1 and Table S1 of
Supplementary Material).
Note that these coordinates have been proposed to be important by
Bolhuis \textit{et al.}~\cite{Bolhuis:2000eo}
The other major components, $\alpha_{57}$, $\alpha_{12}$, and
$\alpha_{11}$, are also the rotations about the
$\mathrm{C-N-C_{\alpha}}$ and $\mathrm{C-N}$ bonds (see
Fig. \ref{fig1}(a)); $\psi$ only comes as a sixth component (as $\alpha_{30}$).
The rotations about $\mathrm{C-N-C_{\alpha}}$ and $\mathrm{C-N}$ bonds,
which can be characterized by $\phi$ and $\theta$, respectively, are
thus suggested to be critical in characterizing the current TS of
interest.

Finally, to confirm this insight, the committor distribution is
examined on the probability distribution of $\phi$ and $\theta$, which
was also obtained from the REMD trajectory and plotted in Fig.~\ref{fig:shooting_point}(a).
Note that the two states A and B are found 
at $\phi \sim -90^{\circ}$ and $\phi \sim 60^{\circ}$, respectively,
whereas the angle $\theta$ is mostly located at $\theta \sim 0^{\circ}$
regardless of the states.
The training dataset points are described as a function of $\phi$ and
$\theta$ in Fig.~\ref{fig:shooting_point}(b).
We see that, in contrast to the $\phi$-$\psi$ plot
in Fig.~\ref{fig:TPS_distribution}(b), the points with $p_\mathrm{B}^* \sim 0.5$ are
narrowly distributed along a diagonal line in the $\phi$-$\theta$ plot
(Fig.~\ref{fig:shooting_point}(b)), indicative of a clearer separatrix.
This confirms that coupled changes of $\phi$ and $\theta$ are important
for the TS along the path connecting states A and B.
It is also consistent with the committor distributions showing
the peak at $p_\mathrm{B}=1/2$ evaluated either by the transition state
sampling~\cite{Bolhuis:2000eo} or by the umbrella
sampling~\cite{Ma:2005jh} on the $\phi$-$\theta$ plane.
In conclusion, it is demonstrated the method of the
minimization of the 
cross-entropy function $\mathcal{H}$ combined with the $L_2$-norm
regularization can guide the straightforward way to 
find the RC that appropriately describes the TS.

\section{Conclusions}
In this paper, we proposed a cross-entropy minimization method to identify the RC from a large
number of CVs using the committor dataset $p_\mathrm{B}^*$.
The method is a generalization of the likelihood maximization approach
proposed by Peters \textit{et al.},~\cite{Peters:2007et} and is also
derived from the Kullback--Leibler divergence.~\cite{Mori:2020fs}
To take account of a large number of CVs and yet avoid overfitting, we
further introduced the $L_2$-norm regularization
technique.~\cite{Bishop:2006ui}

Using the training and test datasets of committor $p_\mathrm{B}^*$, which are
described as a function of the dihedral angles (in the cosine and sine
forms), we minimized the cross-entropy function $\mathcal{H}$ and
determined the optimal balance of the regularization penalty.
We identified the appropriate RC capable of describing the TS of the
isomerization reaction of alanine dipeptide in vacuum.
The minimization of $\mathcal{H}$ was found to be quite stable, \textit{i.e.}, the
parameters consistently converged to the same set independent of the
initial guesses of $\bm{\alpha}$.
The committor distribution at the TS ($r \sim 0$) was found to be
peaked at $p_\mathrm{B} \sim 0.5$, both in the cases of the training and
data sets.
This result indicates that $r = 0$  indeed describes the TS.
The optimized coordinate was dominantly characterized by the dihedral angles
$\phi$ and $\theta$.
These CVs were further justified by the clear separatrix on 
the scattering plot on the $(\phi, \theta)$ plane.
The presented result is consistent with the observation in the previous 
studies~\cite{Bolhuis:2000eo, Ma:2005jh, Ren:2005kb}, 
which showed the importance of $\theta$ in characterizing the TS of this
reaction.

Finally, 
it should be emphasized
that selecting the appropriate RC 
becomes often cumbersome when considered CVs are possibly redundant and
are also correlated with each other.~\cite{Peters:2017tf}
The current approach via the cross-entropy function combined with 
the $L_2$-norm regularization can be a
powerful means to identify and characterize the RC from the $p_\mathrm{B}^*$ dataset.

\section*{Supplementary material}
See supplementary material for dihedral angles and CV indices (Fig.~S1
and Table~S1), full list of optimal coordinate for $\lambda=0$, 0.5, and
10 (Table~S2, Table~S3, and Table~S4, respectively),
committor distributions as a function of the optimized coordinate for
$\lambda=0$ and 10 (Fig.~S2), and 
$p_\mathrm{B}$ probability at about the TS of $r$ ($
 -0.2 \le r \le 0.2$) for
$\lambda=0$ and 10 (Fig.~S3).

\section*{Data availability statement}
The data supporting the findings of this study are available from the
corresponding authors upon reasonable request.


\begin{acknowledgments}
The authors thank Shinji Saito and Takenobu Nakamura for helpful discussions.
This work was partially supported by JSPS KAKENHI Grant Numbers: JP18H02415~(K.O.),
 JP18K05049~(T.M.), JP18H01188~(K.K.), JP20H05221~(K.K.), and
 JP19H04206~(N.M.).
T.M. and K.K. thank the support from the KAKENHI Innovative Area ``Studying the
 Function of Soft Molecular Systems by the Concerted Use of Theory and
 Experiment.'' 
K.O. was supported by Building of Consortia for the Development of Human
 Resources in Science and Technology, MEXT, Japan.
This work was also partially supported by the Fugaku Supercomputing Project
 and the Elements Strategy Initiative for Catalysts and Batteries (No.~JPMXP0112101003) from
 the Ministry of Education, Culture, Sports, Science, and Technology.
The numerical calculations were performed at Research Center of
 Computational Science, Okazaki Research Facilities, National Institutes
 of Natural Sciences, Japan.
\end{acknowledgments}


\begin{thebibliography}{59}%
\makeatletter
\providecommand \@ifxundefined [1]{%
 \@ifx{#1\undefined}
}%
\providecommand \@ifnum [1]{%
 \ifnum #1\expandafter \@firstoftwo
 \else \expandafter \@secondoftwo
 \fi
}%
\providecommand \@ifx [1]{%
 \ifx #1\expandafter \@firstoftwo
 \else \expandafter \@secondoftwo
 \fi
}%
\providecommand \natexlab [1]{#1}%
\providecommand \enquote  [1]{``#1''}%
\providecommand \bibnamefont  [1]{#1}%
\providecommand \bibfnamefont [1]{#1}%
\providecommand \citenamefont [1]{#1}%
\providecommand \href@noop [0]{\@secondoftwo}%
\providecommand \href [0]{\begingroup \@sanitize@url \@href}%
\providecommand \@href[1]{\@@startlink{#1}\@@href}%
\providecommand \@@href[1]{\endgroup#1\@@endlink}%
\providecommand \@sanitize@url [0]{\catcode `\\12\catcode `\$12\catcode
  `\&12\catcode `\#12\catcode `\^12\catcode `\_12\catcode `\%12\relax}%
\providecommand \@@startlink[1]{}%
\providecommand \@@endlink[0]{}%
\providecommand \url  [0]{\begingroup\@sanitize@url \@url }%
\providecommand \@url [1]{\endgroup\@href {#1}{\urlprefix }}%
\providecommand \urlprefix  [0]{URL }%
\providecommand \Eprint [0]{\href }%
\providecommand \doibase [0]{http://dx.doi.org/}%
\providecommand \selectlanguage [0]{\@gobble}%
\providecommand \bibinfo  [0]{\@secondoftwo}%
\providecommand \bibfield  [0]{\@secondoftwo}%
\providecommand \translation [1]{[#1]}%
\providecommand \BibitemOpen [0]{}%
\providecommand \bibitemStop [0]{}%
\providecommand \bibitemNoStop [0]{.\EOS\space}%
\providecommand \EOS [0]{\spacefactor3000\relax}%
\providecommand \BibitemShut  [1]{\csname bibitem#1\endcsname}%
\let\auto@bib@innerbib\@empty
\bibitem [{\citenamefont {Chipot}\ and\ \citenamefont
  {Pohorille}(2007)}]{Chipot:2007wt}%
  \BibitemOpen
  \bibfield  {author} {\bibinfo {author} {\bibfnamefont {C.}~\bibnamefont
  {Chipot}}\ and\ \bibinfo {author} {\bibfnamefont {A.}~\bibnamefont
  {Pohorille}},\ }\href@noop {} {\emph {\bibinfo {title} {{Free Energy
  Calculations: Theory and Applications in Chemistry and Biology}}}}\ (\bibinfo
   {publisher} {Springer},\ \bibinfo {address} {New York},\ \bibinfo {year}
  {2007})\BibitemShut {NoStop}%
\bibitem [{\citenamefont {Zuckerman}(2010)}]{Zuckerman:2010vu}%
  \BibitemOpen
  \bibfield  {author} {\bibinfo {author} {\bibfnamefont {D.~M.}\ \bibnamefont
  {Zuckerman}},\ }\href@noop {} {\emph {\bibinfo {title} {{Statistical Physics
  of Biomolecules: An Introduction }}}}\ (\bibinfo  {publisher} {CRC Press},\
  \bibinfo {address} {Boca Raton},\ \bibinfo {year} {2010})\BibitemShut
  {NoStop}%
\bibitem [{\citenamefont {Torrie}\ and\ \citenamefont
  {Valleau}(1977)}]{Torrie:1977hs}%
  \BibitemOpen
  \bibfield  {author} {\bibinfo {author} {\bibfnamefont {G.~M.}\ \bibnamefont
  {Torrie}}\ and\ \bibinfo {author} {\bibfnamefont {J.~P.}\ \bibnamefont
  {Valleau}},\ }\bibfield  {title} {\enquote {\bibinfo {title} {{Nonphysical
  sampling distributions in Monte Carlo free-energy estimation: Umbrella
  sampling}},}\ }\href@noop {} {\bibfield  {journal} {\bibinfo  {journal} {J.
  Comput. Phys.}\ }\textbf {\bibinfo {volume} {23}},\ \bibinfo {pages}
  {187--199} (\bibinfo {year} {1977})}\BibitemShut {NoStop}%
\bibitem [{\citenamefont {Sugita}\ and\ \citenamefont
  {Okamoto}(1999)}]{Sugita:1999cl}%
  \BibitemOpen
  \bibfield  {author} {\bibinfo {author} {\bibfnamefont {Y.}~\bibnamefont
  {Sugita}}\ and\ \bibinfo {author} {\bibfnamefont {Y.}~\bibnamefont
  {Okamoto}},\ }\bibfield  {title} {\enquote {\bibinfo {title}
  {{Replica-exchange molecular dynamics method for protein folding}},}\
  }\href@noop {} {\bibfield  {journal} {\bibinfo  {journal} {Chem. Phys.
  Lett.}\ }\textbf {\bibinfo {volume} {314}},\ \bibinfo {pages} {141--151}
  (\bibinfo {year} {1999})}\BibitemShut {NoStop}%
\bibitem [{\citenamefont {Laio}\ and\ \citenamefont
  {Parrinello}(2002)}]{Laio:2002ft}%
  \BibitemOpen
  \bibfield  {author} {\bibinfo {author} {\bibfnamefont {A.}~\bibnamefont
  {Laio}}\ and\ \bibinfo {author} {\bibfnamefont {M.}~\bibnamefont
  {Parrinello}},\ }\bibfield  {title} {\enquote {\bibinfo {title} {{Escaping
  free-energy minima}},}\ }\href@noop {} {\bibfield  {journal} {\bibinfo
  {journal} {Proc. Natl. Acad. Sci. U.S.A.}\ }\textbf {\bibinfo {volume}
  {99}},\ \bibinfo {pages} {12562--12566} (\bibinfo {year} {2002})}\BibitemShut
  {NoStop}%
\bibitem [{\citenamefont {Peters}(2017)}]{Peters:2017tf}%
  \BibitemOpen
  \bibfield  {author} {\bibinfo {author} {\bibfnamefont {B.}~\bibnamefont
  {Peters}},\ }\href@noop {} {\emph {\bibinfo {title} {{Reaction Rate Theory
  and Rare Events}}}}\ (\bibinfo  {publisher} {Elsevier},\ \bibinfo {address}
  {Amsterdam},\ \bibinfo {year} {2017})\BibitemShut {NoStop}%
\bibitem [{\citenamefont {Bolhuis}\ \emph {et~al.}(2002)\citenamefont
  {Bolhuis}, \citenamefont {Chandler}, \citenamefont {Dellago},\ and\
  \citenamefont {Geissler}}]{Bolhuis:2002ew}%
  \BibitemOpen
  \bibfield  {author} {\bibinfo {author} {\bibfnamefont {P.~G.}\ \bibnamefont
  {Bolhuis}}, \bibinfo {author} {\bibfnamefont {D.}~\bibnamefont {Chandler}},
  \bibinfo {author} {\bibfnamefont {C.}~\bibnamefont {Dellago}}, \ and\
  \bibinfo {author} {\bibfnamefont {P.~L.}\ \bibnamefont {Geissler}},\
  }\bibfield  {title} {\enquote {\bibinfo {title} {{Transition path sampling:
  throwing ropes over rough mountain passes, in the dark.}}}\ }\href@noop {}
  {\bibfield  {journal} {\bibinfo  {journal} {Annu. Rev. Phys. Chem.}\ }\textbf
  {\bibinfo {volume} {53}},\ \bibinfo {pages} {291--318} (\bibinfo {year}
  {2002})}\BibitemShut {NoStop}%
\bibitem [{\citenamefont {Du}\ \emph {et~al.}(1998)\citenamefont {Du},
  \citenamefont {Pande}, \citenamefont {Grosberg}, \citenamefont {Tanaka},\
  and\ \citenamefont {Shakhnovich}}]{Du:1998iu}%
  \BibitemOpen
  \bibfield  {author} {\bibinfo {author} {\bibfnamefont {R.}~\bibnamefont
  {Du}}, \bibinfo {author} {\bibfnamefont {V.~S.}\ \bibnamefont {Pande}},
  \bibinfo {author} {\bibfnamefont {A.~Y.}\ \bibnamefont {Grosberg}}, \bibinfo
  {author} {\bibfnamefont {T.}~\bibnamefont {Tanaka}}, \ and\ \bibinfo {author}
  {\bibfnamefont {E.~S.}\ \bibnamefont {Shakhnovich}},\ }\bibfield  {title}
  {\enquote {\bibinfo {title} {{On the transition coordinate for protein
  folding}},}\ }\href@noop {} {\bibfield  {journal} {\bibinfo  {journal} {J.
  Chem. Phys.}\ }\textbf {\bibinfo {volume} {108}},\ \bibinfo {pages}
  {334--350} (\bibinfo {year} {1998})}\BibitemShut {NoStop}%
\bibitem [{\citenamefont {Geissler}, \citenamefont {Dellago},\ and\
  \citenamefont {Chandler}(1999)}]{Geissler:1999kj}%
  \BibitemOpen
  \bibfield  {author} {\bibinfo {author} {\bibfnamefont {P.~L.}\ \bibnamefont
  {Geissler}}, \bibinfo {author} {\bibfnamefont {C.}~\bibnamefont {Dellago}}, \
  and\ \bibinfo {author} {\bibfnamefont {D.}~\bibnamefont {Chandler}},\
  }\bibfield  {title} {\enquote {\bibinfo {title} {{Kinetic Pathways of Ion
  Pair Dissociation in Water}},}\ }\href@noop {} {\bibfield  {journal}
  {\bibinfo  {journal} {J. Phys. Chem. B}\ }\textbf {\bibinfo {volume} {103}},\
  \bibinfo {pages} {3706--3710} (\bibinfo {year} {1999})}\BibitemShut {NoStop}%
\bibitem [{\citenamefont {Bolhuis}, \citenamefont {Dellago},\ and\
  \citenamefont {Chandler}(2000)}]{Bolhuis:2000eo}%
  \BibitemOpen
  \bibfield  {author} {\bibinfo {author} {\bibfnamefont {P.~G.}\ \bibnamefont
  {Bolhuis}}, \bibinfo {author} {\bibfnamefont {C.}~\bibnamefont {Dellago}}, \
  and\ \bibinfo {author} {\bibfnamefont {D.}~\bibnamefont {Chandler}},\
  }\bibfield  {title} {\enquote {\bibinfo {title} {{Reaction coordinates of
  biomolecular isomerization}},}\ }\href@noop {} {\bibfield  {journal}
  {\bibinfo  {journal} {Proc. Natl. Acad. Sci. U.S.A.}\ }\textbf {\bibinfo
  {volume} {97}},\ \bibinfo {pages} {5877--5882} (\bibinfo {year}
  {2000})}\BibitemShut {NoStop}%
\bibitem [{\citenamefont {Dellago}, \citenamefont {Bolhuis},\ and\
  \citenamefont {Geissler}(2002)}]{Dellago:2002db}%
  \BibitemOpen
  \bibfield  {author} {\bibinfo {author} {\bibfnamefont {C.}~\bibnamefont
  {Dellago}}, \bibinfo {author} {\bibfnamefont {P.~G.}\ \bibnamefont
  {Bolhuis}}, \ and\ \bibinfo {author} {\bibfnamefont {P.~L.}\ \bibnamefont
  {Geissler}},\ }\bibfield  {title} {\enquote {\bibinfo {title} {{Transition
  Path Sampling}},}\ }in\ \href@noop {} {\emph {\bibinfo {booktitle} {Adv.
  Chem. Phys.}}},\ Vol.\ \bibinfo {volume} {123}\ (\bibinfo  {publisher} {John
  Wiley {\&} Sons, Ltd},\ \bibinfo {year} {2002})\ pp.\ \bibinfo {pages}
  {1--78}\BibitemShut {NoStop}%
\bibitem [{\citenamefont {Hagan}\ \emph {et~al.}(2003)\citenamefont {Hagan},
  \citenamefont {Dinner}, \citenamefont {Chandler},\ and\ \citenamefont
  {Chakraborty}}]{Hagan:2003kt}%
  \BibitemOpen
  \bibfield  {author} {\bibinfo {author} {\bibfnamefont {M.~F.}\ \bibnamefont
  {Hagan}}, \bibinfo {author} {\bibfnamefont {A.~R.}\ \bibnamefont {Dinner}},
  \bibinfo {author} {\bibfnamefont {D.}~\bibnamefont {Chandler}}, \ and\
  \bibinfo {author} {\bibfnamefont {A.~K.}\ \bibnamefont {Chakraborty}},\
  }\bibfield  {title} {\enquote {\bibinfo {title} {{Atomistic understanding of
  kinetic pathways for single base-pair binding and unbinding in DNA}},}\
  }\href@noop {} {\bibfield  {journal} {\bibinfo  {journal} {Proc. Natl. Acad.
  Sci. U.S.A.}\ }\textbf {\bibinfo {volume} {100}},\ \bibinfo {pages}
  {13922--13927} (\bibinfo {year} {2003})}\BibitemShut {NoStop}%
\bibitem [{\citenamefont {Hummer}(2004)}]{Hummer:2004ib}%
  \BibitemOpen
  \bibfield  {author} {\bibinfo {author} {\bibfnamefont {G.}~\bibnamefont
  {Hummer}},\ }\bibfield  {title} {\enquote {\bibinfo {title} {{From transition
  paths to transition states and rate coefficients}},}\ }\href@noop {}
  {\bibfield  {journal} {\bibinfo  {journal} {J. Chem. Phys.}\ }\textbf
  {\bibinfo {volume} {120}},\ \bibinfo {pages} {516--523} (\bibinfo {year}
  {2004})}\BibitemShut {NoStop}%
\bibitem [{\citenamefont {Pan}\ and\ \citenamefont
  {Chandler}(2004)}]{Pan:2004eu}%
  \BibitemOpen
  \bibfield  {author} {\bibinfo {author} {\bibfnamefont {A.~C.}\ \bibnamefont
  {Pan}}\ and\ \bibinfo {author} {\bibfnamefont {D.}~\bibnamefont {Chandler}},\
  }\bibfield  {title} {\enquote {\bibinfo {title} {{Dynamics of Nucleation in
  the Ising Model}},}\ }\href@noop {} {\bibfield  {journal} {\bibinfo
  {journal} {J. Phys. Chem. B}\ }\textbf {\bibinfo {volume} {108}},\ \bibinfo
  {pages} {19681--19686} (\bibinfo {year} {2004})}\BibitemShut {NoStop}%
\bibitem [{\citenamefont {Ma}\ and\ \citenamefont {Dinner}(2005)}]{Ma:2005jh}%
  \BibitemOpen
  \bibfield  {author} {\bibinfo {author} {\bibfnamefont {A.}~\bibnamefont
  {Ma}}\ and\ \bibinfo {author} {\bibfnamefont {A.~R.}\ \bibnamefont
  {Dinner}},\ }\bibfield  {title} {\enquote {\bibinfo {title} {{Automatic
  Method for Identifying Reaction Coordinates in Complex Systems}},}\
  }\href@noop {} {\bibfield  {journal} {\bibinfo  {journal} {J. Phys. Chem. B}\
  }\textbf {\bibinfo {volume} {109}},\ \bibinfo {pages} {6769--6779} (\bibinfo
  {year} {2005})}\BibitemShut {NoStop}%
\bibitem [{\citenamefont {Ren}\ \emph {et~al.}(2005)\citenamefont {Ren},
  \citenamefont {Vanden-Eijnden}, \citenamefont {Maragakis},\ and\
  \citenamefont {E}}]{Ren:2005kb}%
  \BibitemOpen
  \bibfield  {author} {\bibinfo {author} {\bibfnamefont {W.}~\bibnamefont
  {Ren}}, \bibinfo {author} {\bibfnamefont {E.}~\bibnamefont {Vanden-Eijnden}},
  \bibinfo {author} {\bibfnamefont {P.}~\bibnamefont {Maragakis}}, \ and\
  \bibinfo {author} {\bibfnamefont {W.}~\bibnamefont {E}},\ }\bibfield  {title}
  {\enquote {\bibinfo {title} {{Transition pathways in complex systems:
  Application of the finite-temperature string method to the alanine
  dipeptide}},}\ }\href@noop {} {\bibfield  {journal} {\bibinfo  {journal} {J.
  Chem. Phys.}\ }\textbf {\bibinfo {volume} {123}},\ \bibinfo {pages} {134109}
  (\bibinfo {year} {2005})}\BibitemShut {NoStop}%
\bibitem [{\citenamefont {Rhee}\ and\ \citenamefont
  {Pande}(2005)}]{Rhee:2005bz}%
  \BibitemOpen
  \bibfield  {author} {\bibinfo {author} {\bibfnamefont {Y.~M.}\ \bibnamefont
  {Rhee}}\ and\ \bibinfo {author} {\bibfnamefont {V.~S.}\ \bibnamefont
  {Pande}},\ }\bibfield  {title} {\enquote {\bibinfo {title} {{One-Dimensional
  Reaction Coordinate and the Corresponding Potential of Mean Force from
  Commitment Probability Distribution}},}\ }\href@noop {} {\bibfield  {journal}
  {\bibinfo  {journal} {J. Phys. Chem. B}\ }\textbf {\bibinfo {volume} {109}},\
  \bibinfo {pages} {6780--6786} (\bibinfo {year} {2005})}\BibitemShut {NoStop}%
\bibitem [{\citenamefont {E}, \citenamefont {Ren},\ and\ \citenamefont
  {Vanden-Eijnden}(2005)}]{E:2005ho}%
  \BibitemOpen
  \bibfield  {author} {\bibinfo {author} {\bibfnamefont {W.}~\bibnamefont {E}},
  \bibinfo {author} {\bibfnamefont {W.}~\bibnamefont {Ren}}, \ and\ \bibinfo
  {author} {\bibfnamefont {E.}~\bibnamefont {Vanden-Eijnden}},\ }\bibfield
  {title} {\enquote {\bibinfo {title} {{Transition pathways in complex systems:
  Reaction coordinates, isocommittor surfaces, and transition tubes}},}\
  }\href@noop {} {\bibfield  {journal} {\bibinfo  {journal} {Chem. Phys.
  Lett.}\ }\textbf {\bibinfo {volume} {413}},\ \bibinfo {pages} {242--247}
  (\bibinfo {year} {2005})}\BibitemShut {NoStop}%
\bibitem [{\citenamefont {Berezhkovskii}\ and\ \citenamefont
  {Szabo}(2005)}]{Berezhkovskii:2005er}%
  \BibitemOpen
  \bibfield  {author} {\bibinfo {author} {\bibfnamefont {A.}~\bibnamefont
  {Berezhkovskii}}\ and\ \bibinfo {author} {\bibfnamefont {A.}~\bibnamefont
  {Szabo}},\ }\bibfield  {title} {\enquote {\bibinfo {title} {{One-dimensional
  reaction coordinates for diffusive activated rate processes in many
  dimensions}},}\ }\href@noop {} {\bibfield  {journal} {\bibinfo  {journal} {J.
  Chem. Phys.}\ }\textbf {\bibinfo {volume} {122}},\ \bibinfo {pages} {014503}
  (\bibinfo {year} {2005})}\BibitemShut {NoStop}%
\bibitem [{\citenamefont {Best}\ and\ \citenamefont
  {Hummer}(2005)}]{Best:2005cx}%
  \BibitemOpen
  \bibfield  {author} {\bibinfo {author} {\bibfnamefont {R.~B.}\ \bibnamefont
  {Best}}\ and\ \bibinfo {author} {\bibfnamefont {G.}~\bibnamefont {Hummer}},\
  }\bibfield  {title} {\enquote {\bibinfo {title} {{Reaction coordinates and
  rates from transition paths}},}\ }\href@noop {} {\bibfield  {journal}
  {\bibinfo  {journal} {Proc. Natl. Acad. Sci. U.S.A.}\ }\textbf {\bibinfo
  {volume} {102}},\ \bibinfo {pages} {6732--6737} (\bibinfo {year}
  {2005})}\BibitemShut {NoStop}%
\bibitem [{\citenamefont {Moroni}, \citenamefont {ten Wolde},\ and\
  \citenamefont {Bolhuis}(2005)}]{Moroni:2005es}%
  \BibitemOpen
  \bibfield  {author} {\bibinfo {author} {\bibfnamefont {D.}~\bibnamefont
  {Moroni}}, \bibinfo {author} {\bibfnamefont {P.~R.}\ \bibnamefont {ten
  Wolde}}, \ and\ \bibinfo {author} {\bibfnamefont {P.~G.}\ \bibnamefont
  {Bolhuis}},\ }\bibfield  {title} {\enquote {\bibinfo {title} {{Interplay
  between Structure and Size in a Critical Crystal Nucleus}},}\ }\href@noop {}
  {\bibfield  {journal} {\bibinfo  {journal} {Phys. Rev. Lett.}\ }\textbf
  {\bibinfo {volume} {94}},\ \bibinfo {pages} {235703} (\bibinfo {year}
  {2005})}\BibitemShut {NoStop}%
\bibitem [{\citenamefont {Peters}(2006)}]{Peters:2006cv}%
  \BibitemOpen
  \bibfield  {author} {\bibinfo {author} {\bibfnamefont {B.}~\bibnamefont
  {Peters}},\ }\bibfield  {title} {\enquote {\bibinfo {title} {{Using the
  histogram test to quantify reaction coordinate error}},}\ }\href@noop {}
  {\bibfield  {journal} {\bibinfo  {journal} {J. Chem. Phys.}\ }\textbf
  {\bibinfo {volume} {125}},\ \bibinfo {pages} {241101}  (\bibinfo
 {year} {2006})}\BibitemShut {NoStop}%
\bibitem [{\citenamefont {Branduardi}, \citenamefont {Gervasio},\ and\
  \citenamefont {Parrinello}(2007)}]{Branduardi:2007ib}%
  \BibitemOpen
  \bibfield  {author} {\bibinfo {author} {\bibfnamefont {D.}~\bibnamefont
  {Branduardi}}, \bibinfo {author} {\bibfnamefont {F.~L.}\ \bibnamefont
  {Gervasio}}, \ and\ \bibinfo {author} {\bibfnamefont {M.}~\bibnamefont
  {Parrinello}},\ }\bibfield  {title} {\enquote {\bibinfo {title} {{From A to B
  in free energy space}},}\ }\href@noop {} {\bibfield  {journal} {\bibinfo
  {journal} {J. Chem. Phys.}\ }\textbf {\bibinfo {volume} {126}},\ \bibinfo
  {pages} {054103} (\bibinfo {year} {2007})}\BibitemShut {NoStop}%
\bibitem [{\citenamefont {Quaytman}\ and\ \citenamefont
  {Schwartz}(2007)}]{Quaytman:2007vo}%
  \BibitemOpen
  \bibfield  {author} {\bibinfo {author} {\bibfnamefont {S.~L.}\ \bibnamefont
  {Quaytman}}\ and\ \bibinfo {author} {\bibfnamefont {S.~D.}\ \bibnamefont
  {Schwartz}},\ }\bibfield  {title} {\enquote {\bibinfo {title} {{Reaction
  coordinate of an enzymatic reaction revealed by transition path sampling}},}\
  }\href@noop {} {\bibfield  {journal} {\bibinfo  {journal} {Proc. Natl. Acad.
  Sci. U.S.A.}\ }\textbf {\bibinfo {volume} {104}},\ \bibinfo {pages}
  {12253--12258} (\bibinfo {year} {2007})}\BibitemShut {NoStop}%
\bibitem [{\citenamefont {Antoniou}\ and\ \citenamefont
  {Schwartz}(2009)}]{Antoniou:2009gg}%
  \BibitemOpen
  \bibfield  {author} {\bibinfo {author} {\bibfnamefont {D.}~\bibnamefont
  {Antoniou}}\ and\ \bibinfo {author} {\bibfnamefont {S.~D.}\ \bibnamefont
  {Schwartz}},\ }\bibfield  {title} {\enquote {\bibinfo {title} {{The
  stochastic separatrix and the reaction coordinate for complex systems}},}\
  }\href@noop {} {\bibfield  {journal} {\bibinfo  {journal} {J. Chem. Phys.}\
  }\textbf {\bibinfo {volume} {130}},\ \bibinfo {pages} {151103} (\bibinfo
  {year} {2009})}\BibitemShut {NoStop}%
\bibitem [{\citenamefont {Peters}(2010{\natexlab{a}})}]{Peters:2010ku}%
  \BibitemOpen
  \bibfield  {author} {\bibinfo {author} {\bibfnamefont {B.}~\bibnamefont
  {Peters}},\ }\bibfield  {title} {\enquote {\bibinfo {title} {{p(TP\textbar q)
  peak maximization: Necessary but not sufficient for reaction coordinate
  accuracy}},}\ }\href@noop {} {\bibfield  {journal} {\bibinfo  {journal}
  {Chem. Phys. Lett.}\ }\textbf {\bibinfo {volume} {494}},\ \bibinfo {pages}
  {100--103} (\bibinfo {year} {2010}{\natexlab{a}})}\BibitemShut {NoStop}%
\bibitem [{\citenamefont {Peters}(2010{\natexlab{b}})}]{Peters:2010hm}%
  \BibitemOpen
  \bibfield  {author} {\bibinfo {author} {\bibfnamefont {B.}~\bibnamefont
  {Peters}},\ }\bibfield  {title} {\enquote {\bibinfo {title} {{Recent advances
  in transition path sampling: accurate reaction coordinates, likelihood
  maximisation and diffusive barrier-crossing dynamics}},}\ }\href@noop {}
  {\bibfield  {journal} {\bibinfo  {journal} {Mol. Simul.}\ }\textbf {\bibinfo
  {volume} {36}},\ \bibinfo {pages} {1265--1281} (\bibinfo {year}
  {2010}{\natexlab{b}})}\BibitemShut {NoStop}%
\bibitem [{\citenamefont {Li}\ and\ \citenamefont {Ma}(2014)}]{Li:2014fp}%
  \BibitemOpen
  \bibfield  {author} {\bibinfo {author} {\bibfnamefont {W.}~\bibnamefont
  {Li}}\ and\ \bibinfo {author} {\bibfnamefont {A.}~\bibnamefont {Ma}},\
  }\bibfield  {title} {\enquote {\bibinfo {title} {{Recent developments in
  methods for identifying reaction coordinates.}}}\ }\href@noop {} {\bibfield
  {journal} {\bibinfo  {journal} {Mol. Simul.}\ }\textbf {\bibinfo {volume}
  {40}},\ \bibinfo {pages} {784--793} (\bibinfo {year} {2014})}\BibitemShut
  {NoStop}%
\bibitem [{\citenamefont {Wales}(2015)}]{Wales:2015dl}%
  \BibitemOpen
  \bibfield  {author} {\bibinfo {author} {\bibfnamefont {D.~J.}\ \bibnamefont
  {Wales}},\ }\bibfield  {title} {\enquote {\bibinfo {title} {{Perspective:
  Insight into reaction coordinates and dynamics from the potential energy
  landscape}},}\ }\href@noop {} {\bibfield  {journal} {\bibinfo  {journal} {J.
  Chem. Phys.}\ }\textbf {\bibinfo {volume} {142}},\ \bibinfo {pages} {130901}
  (\bibinfo {year} {2015})}\BibitemShut {NoStop}%
\bibitem [{\citenamefont {Peters}(2016)}]{Peters:2016by}%
  \BibitemOpen
  \bibfield  {author} {\bibinfo {author} {\bibfnamefont {B.}~\bibnamefont
  {Peters}},\ }\bibfield  {title} {\enquote {\bibinfo {title} {{Reaction
  Coordinates and Mechanistic Hypothesis Tests}},}\ }\href@noop {} {\bibfield
  {journal} {\bibinfo  {journal} {Annu. Rev. Phys. Chem.}\ }\textbf {\bibinfo
  {volume} {67}},\ \bibinfo {pages} {669--690} (\bibinfo {year}
  {2016})}\BibitemShut {NoStop}%
\bibitem [{\citenamefont {Banushkina}\ and\ \citenamefont
  {Krivov}(2016)}]{Banushkina:2016hi}%
  \BibitemOpen
  \bibfield  {author} {\bibinfo {author} {\bibfnamefont {P.~V.}\ \bibnamefont
  {Banushkina}}\ and\ \bibinfo {author} {\bibfnamefont {S.~V.}\ \bibnamefont
  {Krivov}},\ }\bibfield  {title} {\enquote {\bibinfo {title} {{Optimal
  reaction coordinates}},}\ }\href@noop {} {\bibfield  {journal} {\bibinfo
  {journal} {WIREs Comput Mol Sci}\ }\textbf {\bibinfo {volume} {6}},\ \bibinfo
  {pages} {748--763} (\bibinfo {year} {2016})}\BibitemShut {NoStop}%
\bibitem [{\citenamefont {Sittel}\ and\ \citenamefont
  {Stock}(2018)}]{Sittel:2018gs}%
  \BibitemOpen
  \bibfield  {author} {\bibinfo {author} {\bibfnamefont {F.}~\bibnamefont
  {Sittel}}\ and\ \bibinfo {author} {\bibfnamefont {G.}~\bibnamefont {Stock}},\
  }\bibfield  {title} {\enquote {\bibinfo {title} {{Perspective: Identification
  of collective variables and metastable states of protein dynamics}},}\
  }\href@noop {} {\bibfield  {journal} {\bibinfo  {journal} {J. Chem. Phys.}\
  }\textbf {\bibinfo {volume} {149}},\ \bibinfo {pages} {150901} (\bibinfo
  {year} {2018})}\BibitemShut {NoStop}%
\bibitem [{\citenamefont {Sultan}\ and\ \citenamefont
  {Pande}(2018)}]{Sultan:2018fo}%
  \BibitemOpen
  \bibfield  {author} {\bibinfo {author} {\bibfnamefont {M.~M.}\ \bibnamefont
  {Sultan}}\ and\ \bibinfo {author} {\bibfnamefont {V.~S.}\ \bibnamefont
  {Pande}},\ }\bibfield  {title} {\enquote {\bibinfo {title} {{Automated design
  of collective variables using supervised machine learning}},}\ }\href@noop {}
  {\bibfield  {journal} {\bibinfo  {journal} {J. Chem. Phys.}\ }\textbf
  {\bibinfo {volume} {149}},\ \bibinfo {pages} {094106} (\bibinfo {year}
  {2018})}\BibitemShut {NoStop}%
\bibitem [{\citenamefont {Jung}, \citenamefont {Covino},\ and\ \citenamefont
  {Hummer}(2019)}]{Jung:2019td}%
  \BibitemOpen
  \bibfield  {author} {\bibinfo {author} {\bibfnamefont {H.}~\bibnamefont
  {Jung}}, \bibinfo {author} {\bibfnamefont {R.}~\bibnamefont {Covino}}, \ and\
  \bibinfo {author} {\bibfnamefont {G.}~\bibnamefont {Hummer}},\ }\bibfield
  {title} {\enquote {\bibinfo {title} {{Artificial Intelligence Assists
  Discovery of Reaction Coordinates and Mechanisms from Molecular Dynamics
  Simulations}},}\ }\href@noop {} {\bibfield  {journal} {\bibinfo  {journal}
  {arXiv}\ } (\bibinfo {year} {2019})},\ \Eprint
  {http://arxiv.org/abs/1901.04595v1} {1901.04595v1} \BibitemShut {NoStop}%
\bibitem [{\citenamefont {No{\'e}}\ \emph {et~al.}(2020)\citenamefont
  {No{\'e}}, \citenamefont {Tkatchenko}, \citenamefont {M{\"u}ller},\ and\
  \citenamefont {Clementi}}]{Noe:2020ic}%
  \BibitemOpen
  \bibfield  {author} {\bibinfo {author} {\bibfnamefont {F.}~\bibnamefont
  {No{\'e}}}, \bibinfo {author} {\bibfnamefont {A.}~\bibnamefont {Tkatchenko}},
  \bibinfo {author} {\bibfnamefont {K.-R.}\ \bibnamefont {M{\"u}ller}}, \ and\
  \bibinfo {author} {\bibfnamefont {C.}~\bibnamefont {Clementi}},\ }\bibfield
  {title} {\enquote {\bibinfo {title} {{Machine Learning for Molecular
  Simulation}},}\ }\href@noop {} {\bibfield  {journal} {\bibinfo  {journal}
  {Annu. Rev. Phys. Chem.}\ }\textbf {\bibinfo {volume} {71}},\ \bibinfo
	  {pages} {361--390}
 (\bibinfo {year} {2020})}\BibitemShut {NoStop}%
\bibitem [{\citenamefont {Sidky}, \citenamefont {Chen},\ and\ \citenamefont
  {Ferguson}(2020)}]{Sidky:2020im}%
  \BibitemOpen
  \bibfield  {author} {\bibinfo {author} {\bibfnamefont {H.}~\bibnamefont
  {Sidky}}, \bibinfo {author} {\bibfnamefont {W.}~\bibnamefont {Chen}}, \ and\
  \bibinfo {author} {\bibfnamefont {A.~L.}\ \bibnamefont {Ferguson}},\
  }\bibfield  {title} {\enquote {\bibinfo {title} {{Machine learning for
  collective variable discovery and enhanced sampling in biomolecular
  simulation}},}\ }\href@noop {} {\bibfield  {journal} {\bibinfo  {journal}
  {Mol. Phys.}\ }\textbf {\bibinfo {volume} {118}},\ \bibinfo {pages} {e1737742}
  (\bibinfo {year} {2020})}\BibitemShut {NoStop}%
\bibitem [{\citenamefont {Peters}, \citenamefont {Beckham},\ and\ \citenamefont
  {Trout}(2007)}]{Peters:2007et}%
  \BibitemOpen
  \bibfield  {author} {\bibinfo {author} {\bibfnamefont {B.}~\bibnamefont
  {Peters}}, \bibinfo {author} {\bibfnamefont {G.~T.}\ \bibnamefont {Beckham}},
  \ and\ \bibinfo {author} {\bibfnamefont {B.~L.}\ \bibnamefont {Trout}},\
  }\bibfield  {title} {\enquote {\bibinfo {title} {{Extensions to the
  likelihood maximization approach for finding reaction coordinates}},}\
  }\href@noop {} {\bibfield  {journal} {\bibinfo  {journal} {J. Chem. Phys.}\
  }\textbf {\bibinfo {volume} {127}},\ \bibinfo {pages} {034109} (\bibinfo
  {year} {2007})}\BibitemShut {NoStop}%
\bibitem [{\citenamefont {Peters}\ and\ \citenamefont
  {Trout}(2006)}]{Peters:2006iz}%
  \BibitemOpen
  \bibfield  {author} {\bibinfo {author} {\bibfnamefont {B.}~\bibnamefont
  {Peters}}\ and\ \bibinfo {author} {\bibfnamefont {B.~L.}\ \bibnamefont
  {Trout}},\ }\bibfield  {title} {\enquote {\bibinfo {title} {{Obtaining
  reaction coordinates by likelihood maximization}},}\ }\href@noop {}
  {\bibfield  {journal} {\bibinfo  {journal} {J. Chem. Phys.}\ }\textbf
  {\bibinfo {volume} {125}},\ \bibinfo {pages} {054108} (\bibinfo {year}
  {2006})}\BibitemShut {NoStop}%
\bibitem [{\citenamefont {Beckham}\ \emph {et~al.}(2007)\citenamefont
  {Beckham}, \citenamefont {Peters}, \citenamefont {Starbuck}, \citenamefont
  {Variankaval},\ and\ \citenamefont {Trout}}]{Beckham:2007jz}%
  \BibitemOpen
  \bibfield  {author} {\bibinfo {author} {\bibfnamefont {G.~T.}\ \bibnamefont
  {Beckham}}, \bibinfo {author} {\bibfnamefont {B.}~\bibnamefont {Peters}},
  \bibinfo {author} {\bibfnamefont {C.}~\bibnamefont {Starbuck}}, \bibinfo
  {author} {\bibfnamefont {N.}~\bibnamefont {Variankaval}}, \ and\ \bibinfo
  {author} {\bibfnamefont {B.~L.}\ \bibnamefont {Trout}},\ }\bibfield  {title}
  {\enquote {\bibinfo {title} {{Surface-Mediated Nucleation in the Solid-State
  Polymorph Transformation of Terephthalic Acid}},}\ }\href@noop {} {\bibfield
  {journal} {\bibinfo  {journal} {J. Am. Chem. Soc.}\ }\textbf {\bibinfo
  {volume} {129}},\ \bibinfo {pages} {4714--4723} (\bibinfo {year}
  {2007})}\BibitemShut {NoStop}%
\bibitem [{\citenamefont {Beckham}, \citenamefont {Peters},\ and\ \citenamefont
  {Trout}(2008)}]{Beckham:2008hh}%
  \BibitemOpen
  \bibfield  {author} {\bibinfo {author} {\bibfnamefont {G.~T.}\ \bibnamefont
  {Beckham}}, \bibinfo {author} {\bibfnamefont {B.}~\bibnamefont {Peters}}, \
  and\ \bibinfo {author} {\bibfnamefont {B.~L.}\ \bibnamefont {Trout}},\
  }\bibfield  {title} {\enquote {\bibinfo {title} {{Evidence for a Size
  Dependent Nucleation Mechanism in Solid State Polymorph Transformations}},}\
  }\href@noop {} {\bibfield  {journal} {\bibinfo  {journal} {J. Phys. Chem. B}\
  }\textbf {\bibinfo {volume} {112}},\ \bibinfo {pages} {7460--7466} (\bibinfo
  {year} {2008})}\BibitemShut {NoStop}%
\bibitem [{\citenamefont {Vreede}, \citenamefont {Juraszek},\ and\
  \citenamefont {Bolhuis}(2010)}]{Vreede:2010ig}%
  \BibitemOpen
  \bibfield  {author} {\bibinfo {author} {\bibfnamefont {J.}~\bibnamefont
  {Vreede}}, \bibinfo {author} {\bibfnamefont {J.}~\bibnamefont {Juraszek}}, \
  and\ \bibinfo {author} {\bibfnamefont {P.~G.}\ \bibnamefont {Bolhuis}},\
  }\bibfield  {title} {\enquote {\bibinfo {title} {{Predicting the reaction
  coordinates of millisecond light-induced conformational changes in
  photoactive yellow protein}},}\ }\href@noop {} {\bibfield  {journal}
  {\bibinfo  {journal} {Proc. Natl. Acad. Sci. U.S.A.}\ }\textbf {\bibinfo
  {volume} {107}},\ \bibinfo {pages} {2397--2402} (\bibinfo {year}
  {2010})}\BibitemShut {NoStop}%
\bibitem [{\citenamefont {Lechner}\ \emph {et~al.}(2010)\citenamefont
  {Lechner}, \citenamefont {Rogal}, \citenamefont {Juraszek}, \citenamefont
  {Ensing},\ and\ \citenamefont {Bolhuis}}]{Lechner:2010du}%
  \BibitemOpen
  \bibfield  {author} {\bibinfo {author} {\bibfnamefont {W.}~\bibnamefont
  {Lechner}}, \bibinfo {author} {\bibfnamefont {J.}~\bibnamefont {Rogal}},
  \bibinfo {author} {\bibfnamefont {J.}~\bibnamefont {Juraszek}}, \bibinfo
  {author} {\bibfnamefont {B.}~\bibnamefont {Ensing}}, \ and\ \bibinfo {author}
  {\bibfnamefont {P.~G.}\ \bibnamefont {Bolhuis}},\ }\bibfield  {title}
  {\enquote {\bibinfo {title} {{Nonlinear reaction coordinate analysis in the
  reweighted path ensemble}},}\ }\href@noop {} {\bibfield  {journal} {\bibinfo
  {journal} {J. Chem. Phys.}\ }\textbf {\bibinfo {volume} {133}},\ \bibinfo
  {pages} {174110} (\bibinfo {year} {2010})}\BibitemShut {NoStop}%
\bibitem [{\citenamefont {Pan}\ and\ \citenamefont {Ricci}(2010)}]{Pan:2010ky}%
  \BibitemOpen
  \bibfield  {author} {\bibinfo {author} {\bibfnamefont {B.}~\bibnamefont
  {Pan}}\ and\ \bibinfo {author} {\bibfnamefont {M.~S.}\ \bibnamefont
  {Ricci}},\ }\bibfield  {title} {\enquote {\bibinfo {title} {{Molecular
  Mechanism of Acid-Catalyzed Hydrolysis of Peptide Bonds Using a Model
  Compound}},}\ }\href@noop {} {\bibfield  {journal} {\bibinfo  {journal} {J.
  Phys. Chem. B}\ }\textbf {\bibinfo {volume} {114}},\ \bibinfo {pages}
  {4389--4399} (\bibinfo {year} {2010})}\BibitemShut {NoStop}%
\bibitem [{\citenamefont {Beckham}\ and\ \citenamefont
  {Peters}(2011)}]{Beckham:2011dw}%
  \BibitemOpen
  \bibfield  {author} {\bibinfo {author} {\bibfnamefont {G.~T.}\ \bibnamefont
  {Beckham}}\ and\ \bibinfo {author} {\bibfnamefont {B.}~\bibnamefont
  {Peters}},\ }\bibfield  {title} {\enquote {\bibinfo {title} {{Optimizing
  Nucleus Size Metrics for Liquid--Solid Nucleation from Transition Paths of
  Near-Nanosecond Duration}},}\ }\href@noop {} {\bibfield  {journal} {\bibinfo
  {journal} {J. Phys. Chem. Lett.}\ }\textbf {\bibinfo {volume} {2}},\ \bibinfo
  {pages} {1133--1138} (\bibinfo {year} {2011})}\BibitemShut {NoStop}%
\bibitem [{\citenamefont {Lechner}, \citenamefont {Dellago},\ and\
  \citenamefont {Bolhuis}(2011)}]{Lechner:2011cv}%
  \BibitemOpen
  \bibfield  {author} {\bibinfo {author} {\bibfnamefont {W.}~\bibnamefont
  {Lechner}}, \bibinfo {author} {\bibfnamefont {C.}~\bibnamefont {Dellago}}, \
  and\ \bibinfo {author} {\bibfnamefont {P.~G.}\ \bibnamefont {Bolhuis}},\
  }\bibfield  {title} {\enquote {\bibinfo {title} {{Role of the Prestructured
  Surface Cloud in Crystal Nucleation}},}\ }\href@noop {} {\bibfield  {journal}
  {\bibinfo  {journal} {Phys. Rev. Lett.}\ }\textbf {\bibinfo {volume} {106}},\
  \bibinfo {pages} {085701} (\bibinfo {year} {2011})}\BibitemShut {NoStop}%
\bibitem [{\citenamefont {Peters}(2012)}]{Peters:2012cv}%
  \BibitemOpen
  \bibfield  {author} {\bibinfo {author} {\bibfnamefont {B.}~\bibnamefont
  {Peters}},\ }\bibfield  {title} {\enquote {\bibinfo {title} {{Inertial
  likelihood maximization for reaction coordinates with high transmission
  coefficients}},}\ }\href@noop {} {\bibfield  {journal} {\bibinfo  {journal}
  {Chem. Phys. Lett.}\ }\textbf {\bibinfo {volume} {554}},\ \bibinfo {pages}
  {248--253} (\bibinfo {year} {2012})}\BibitemShut {NoStop}%
\bibitem [{\citenamefont {Xi}, \citenamefont {Shah},\ and\ \citenamefont
  {Trout}(2013)}]{Xi:2013en}%
  \BibitemOpen
  \bibfield  {author} {\bibinfo {author} {\bibfnamefont {L.}~\bibnamefont
  {Xi}}, \bibinfo {author} {\bibfnamefont {M.}~\bibnamefont {Shah}}, \ and\
  \bibinfo {author} {\bibfnamefont {B.~L.}\ \bibnamefont {Trout}},\ }\bibfield
  {title} {\enquote {\bibinfo {title} {{Hopping of Water in a Glassy Polymer
  Studied via Transition Path Sampling and Likelihood Maximization}},}\
  }\href@noop {} {\bibfield  {journal} {\bibinfo  {journal} {J. Phys. Chem. B}\
  }\textbf {\bibinfo {volume} {117}},\ \bibinfo {pages} {3634--3647} (\bibinfo
  {year} {2013})}\BibitemShut {NoStop}%
\bibitem [{\citenamefont {Jungblut}, \citenamefont {Singraber},\ and\
  \citenamefont {Dellago}(2013)}]{Jungblut:2013hn}%
  \BibitemOpen
  \bibfield  {author} {\bibinfo {author} {\bibfnamefont {S.}~\bibnamefont
  {Jungblut}}, \bibinfo {author} {\bibfnamefont {A.}~\bibnamefont {Singraber}},
  \ and\ \bibinfo {author} {\bibfnamefont {C.}~\bibnamefont {Dellago}},\
  }\bibfield  {title} {\enquote {\bibinfo {title} {{Optimising reaction
  coordinates for crystallisation by tuning the crystallinity definition}},}\
  }\href@noop {} {\bibfield  {journal} {\bibinfo  {journal} {Mol. Phys.}\
  }\textbf {\bibinfo {volume} {111}},\ \bibinfo {pages} {3527--3533} (\bibinfo
  {year} {2013})}\BibitemShut {NoStop}%
\bibitem [{\citenamefont {Mullen}, \citenamefont {Shea},\ and\ \citenamefont
  {Peters}(2014)}]{Mullen:2014cl}%
  \BibitemOpen
  \bibfield  {author} {\bibinfo {author} {\bibfnamefont {R.~G.}\ \bibnamefont
  {Mullen}}, \bibinfo {author} {\bibfnamefont {J.-E.}\ \bibnamefont {Shea}}, \
  and\ \bibinfo {author} {\bibfnamefont {B.}~\bibnamefont {Peters}},\
  }\bibfield  {title} {\enquote {\bibinfo {title} {{Transmission Coefficients,
  Committors, and Solvent Coordinates in Ion-Pair Dissociation}},}\ }\href@noop
  {} {\bibfield  {journal} {\bibinfo  {journal} {J. Chem. Theory Comput.}\
  }\textbf {\bibinfo {volume} {10}},\ \bibinfo {pages} {659--667} (\bibinfo
  {year} {2014})}\BibitemShut {NoStop}%
\bibitem [{\citenamefont {Mullen}, \citenamefont {Shea},\ and\ \citenamefont
  {Peters}(2015)}]{Mullen:2015jb}%
  \BibitemOpen
  \bibfield  {author} {\bibinfo {author} {\bibfnamefont {R.~G.}\ \bibnamefont
  {Mullen}}, \bibinfo {author} {\bibfnamefont {J.-E.}\ \bibnamefont {Shea}}, \
  and\ \bibinfo {author} {\bibfnamefont {B.}~\bibnamefont {Peters}},\
  }\bibfield  {title} {\enquote {\bibinfo {title} {{Easy Transition Path
  Sampling Methods: Flexible-Length Aimless Shooting and Permutation
  Shooting}},}\ }\href@noop {} {\bibfield  {journal} {\bibinfo  {journal} {J.
  Chem. Theory Comput.}\ }\textbf {\bibinfo {volume} {11}},\ \bibinfo {pages}
  {2421--2428} (\bibinfo {year} {2015})}\BibitemShut {NoStop}%
\bibitem [{\citenamefont {Lupi}, \citenamefont {Peters},\ and\ \citenamefont
  {Molinero}(2016)}]{Lupi:2016dda}%
  \BibitemOpen
  \bibfield  {author} {\bibinfo {author} {\bibfnamefont {L.}~\bibnamefont
  {Lupi}}, \bibinfo {author} {\bibfnamefont {B.}~\bibnamefont {Peters}}, \ and\
  \bibinfo {author} {\bibfnamefont {V.}~\bibnamefont {Molinero}},\ }\bibfield
  {title} {\enquote {\bibinfo {title} {{Pre-ordering of interfacial water in
  the pathway of heterogeneous ice nucleation does not lead to a two-step
  crystallization mechanism}},}\ }\href@noop {} {\bibfield  {journal} {\bibinfo
   {journal} {J. Chem. Phys.}\ }\textbf {\bibinfo {volume} {145}},\ \bibinfo
  {pages} {211910} (\bibinfo {year} {2016})}\BibitemShut {NoStop}%
\bibitem [{\citenamefont {Jung}, \citenamefont {Okazaki},\ and\ \citenamefont
  {Hummer}(2017)}]{Jung:2017cd}%
  \BibitemOpen
  \bibfield  {author} {\bibinfo {author} {\bibfnamefont {H.}~\bibnamefont
  {Jung}}, \bibinfo {author} {\bibfnamefont {K.-i.}\ \bibnamefont {Okazaki}}, \
  and\ \bibinfo {author} {\bibfnamefont {G.}~\bibnamefont {Hummer}},\
  }\bibfield  {title} {\enquote {\bibinfo {title} {{Transition path sampling of
  rare events by shooting from the top}},}\ }\href@noop {} {\bibfield
  {journal} {\bibinfo  {journal} {J. Chem. Phys.}\ }\textbf {\bibinfo {volume}
  {147}},\ \bibinfo {pages} {152716} (\bibinfo {year} {2017})}\BibitemShut
  {NoStop}%
\bibitem [{\citenamefont {Joswiak}, \citenamefont {Doherty},\ and\
  \citenamefont {Peters}(2018)}]{Joswiak:2018jl}%
  \BibitemOpen
  \bibfield  {author} {\bibinfo {author} {\bibfnamefont {M.~N.}\ \bibnamefont
  {Joswiak}}, \bibinfo {author} {\bibfnamefont {M.~F.}\ \bibnamefont
  {Doherty}}, \ and\ \bibinfo {author} {\bibfnamefont {B.}~\bibnamefont
  {Peters}},\ }\bibfield  {title} {\enquote {\bibinfo {title} {{Ion dissolution
  mechanism and kinetics at kink sites on NaCl surfaces}},}\ }\href@noop {}
  {\bibfield  {journal} {\bibinfo  {journal} {Proc. Natl. Acad. Sci. U.S.A.}\
  }\textbf {\bibinfo {volume} {115}},\ \bibinfo {pages} {656-661} (\bibinfo
  {year} {2018})}\BibitemShut {NoStop}%
\bibitem [{\citenamefont {D{\'\i}az~Leines}\ and\ \citenamefont
  {Rogal}(2018)}]{DiazLeines:2018dh}%
  \BibitemOpen
  \bibfield  {author} {\bibinfo {author} {\bibfnamefont {G.}~\bibnamefont
  {D{\'\i}az~Leines}}\ and\ \bibinfo {author} {\bibfnamefont {J.}~\bibnamefont
  {Rogal}},\ }\bibfield  {title} {\enquote {\bibinfo {title} {{Maximum
  Likelihood Analysis of Reaction Coordinates during Solidification in Ni}},}\
  }\href@noop {} {\bibfield  {journal} {\bibinfo  {journal} {J. Phys. Chem. B}\
  }\textbf {\bibinfo {volume} {122}},\ \bibinfo {pages} {10934--10942}
  (\bibinfo {year} {2018})}\BibitemShut {NoStop}%
\bibitem [{\citenamefont {Okazaki}\ \emph {et~al.}(2019)\citenamefont
  {Okazaki}, \citenamefont {W{\"o}hlert}, \citenamefont {Warnau}, \citenamefont
  {Jung}, \citenamefont {Yildiz}, \citenamefont {K{\"u}hlbrandt},\ and\
  \citenamefont {Hummer}}]{Okazaki:2019jx}%
  \BibitemOpen
  \bibfield  {author} {\bibinfo {author} {\bibfnamefont {K.-i.}\ \bibnamefont
  {Okazaki}}, \bibinfo {author} {\bibfnamefont {D.}~\bibnamefont
  {W{\"o}hlert}}, \bibinfo {author} {\bibfnamefont {J.}~\bibnamefont {Warnau}},
  \bibinfo {author} {\bibfnamefont {H.}~\bibnamefont {Jung}}, \bibinfo {author}
  {\bibfnamefont {{\"O}.}~\bibnamefont {Yildiz}}, \bibinfo {author}
  {\bibfnamefont {W.}~\bibnamefont {K{\"u}hlbrandt}}, \ and\ \bibinfo {author}
  {\bibfnamefont {G.}~\bibnamefont {Hummer}},\ }\bibfield  {title} {\enquote
  {\bibinfo {title} {{Mechanism of the electroneutral sodium/proton antiporter
  PaNhaP from transition-path shooting}},}\ }\href@noop {} {\bibfield
  {journal} {\bibinfo  {journal} {Nat. Commun.}\ }\textbf {\bibinfo {volume}
  {10}},\ \bibinfo {pages} {87} (\bibinfo {year} {2019})}\BibitemShut {NoStop}%
\bibitem [{\citenamefont {Bishop}(2006)}]{Bishop:2006ui}%
  \BibitemOpen
  \bibfield  {author} {\bibinfo {author} {\bibfnamefont {C.}~\bibnamefont
  {Bishop}},\ }\href@noop {} {\emph {\bibinfo {title} {{Pattern Recognition and
  Machine Learning}}}}\ (\bibinfo  {publisher} {Springer},\ \bibinfo {address}
  {New York},\ \bibinfo {year} {2006})\BibitemShut {NoStop}%
\bibitem [{\citenamefont {Mori}\ and\ \citenamefont
  {Saito}(2020)}]{Mori:2020fs}%
  \BibitemOpen
  \bibfield  {author} {\bibinfo {author} {\bibfnamefont {T.}~\bibnamefont
  {Mori}}\ and\ \bibinfo {author} {\bibfnamefont {S.}~\bibnamefont {Saito}},\
  }\bibfield  {title} {\enquote {\bibinfo {title} {{Dissecting the dynamics
  during enzyme catalysis: A case study of Pin1 peptidyl-prolyl isomerase}},}\
  }\href@noop {} {\bibfield  {journal} {\bibinfo  {journal} {J. Chem. Theory
  Comput.}\ }\textbf {\bibinfo {volume} {16}},\ \bibinfo {pages} {3396--4307}
  (\bibinfo {year} {2020})}\BibitemShut {NoStop}%
\bibitem [{\citenamefont {Hornak}\ \emph {et~al.}(2006)\citenamefont {Hornak},
  \citenamefont {Abel}, \citenamefont {Okur}, \citenamefont {Strockbine},
  \citenamefont {Roitberg},\ and\ \citenamefont {Simmerling}}]{Hornak:2006gx}%
  \BibitemOpen
  \bibfield  {author} {\bibinfo {author} {\bibfnamefont {V.}~\bibnamefont
  {Hornak}}, \bibinfo {author} {\bibfnamefont {R.}~\bibnamefont {Abel}},
  \bibinfo {author} {\bibfnamefont {A.}~\bibnamefont {Okur}}, \bibinfo {author}
  {\bibfnamefont {B.}~\bibnamefont {Strockbine}}, \bibinfo {author}
  {\bibfnamefont {A.}~\bibnamefont {Roitberg}}, \ and\ \bibinfo {author}
  {\bibfnamefont {C.}~\bibnamefont {Simmerling}},\ }\bibfield  {title}
  {\enquote {\bibinfo {title} {{Comparison of multiple Amber force fields and
  development of improved protein backbone parameters}},}\ }\href@noop {}
  {\bibfield  {journal} {\bibinfo  {journal} {Proteins}\ }\textbf {\bibinfo
  {volume} {65}},\ \bibinfo {pages} {712--725} (\bibinfo {year}
  {2006})}\BibitemShut {NoStop}%
\bibitem [{\citenamefont {Abraham}\ \emph {et~al.}(2015)\citenamefont
  {Abraham}, \citenamefont {Murtola}, \citenamefont {Schulz}, \citenamefont
  {P{\'a}ll}, \citenamefont {Smith}, \citenamefont {Hess},\ and\ \citenamefont
  {Lindahl}}]{Abraham:2015gj}%
  \BibitemOpen
  \bibfield  {author} {\bibinfo {author} {\bibfnamefont {M.~J.}\ \bibnamefont
  {Abraham}}, \bibinfo {author} {\bibfnamefont {T.}~\bibnamefont {Murtola}},
  \bibinfo {author} {\bibfnamefont {R.}~\bibnamefont {Schulz}}, \bibinfo
  {author} {\bibfnamefont {S.}~\bibnamefont {P{\'a}ll}}, \bibinfo {author}
  {\bibfnamefont {J.~C.}\ \bibnamefont {Smith}}, \bibinfo {author}
  {\bibfnamefont {B.}~\bibnamefont {Hess}}, \ and\ \bibinfo {author}
  {\bibfnamefont {E.}~\bibnamefont {Lindahl}},\ }\bibfield  {title} {\enquote
  {\bibinfo {title} {{GROMACS: High performance molecular simulations through
  multi-level parallelism from laptops to supercomputers}},}\ }\href@noop {}
  {\bibfield  {journal} {\bibinfo  {journal} {SoftwareX}\ }\textbf {\bibinfo
  {volume} {1-2}},\ \bibinfo {pages} {19--25} (\bibinfo {year}
  {2015})}\BibitemShut {NoStop}%
\end{thebibliography}
%

\clearpage
\widetext

\setcounter{equation}{0}
\setcounter{figure}{0}
\setcounter{table}{0}
\setcounter{page}{1}

\noindent{\bf\Large Supplementary Material}
\vspace{5mm}
\begin{center}
\textbf{\large  Learning reaction coordinates via cross-entropy minimization:\\ Application to alanine dipeptide}
\\

\vspace{5mm}
 
{Yusuke Mori$^1$, Kei-ichi Okazaki$^2$, Toshifumi Mori$^{2, 3}$, Kang Kim$^{1, 2}$,
 Nobuyuki Matubayasi$^{1}$}
\\

\vspace{5mm}

\noindent
\textit{
$^1$Division of Chemical Engineering, Graduate School of Engineering Science, Osaka University, Osaka 560-8531, Japan\\
$^2$Institute for Molecular Science, Okazaki, Aichi 444-8585, Japan\\
$^3$The Graduate University for Advanced Studies, Okazaki, Aichi 444-8585, Japan}

\end{center}

\setcounter{equation}{0}
\renewcommand{\theequation}{S.\arabic{equation}}
\renewcommand{\thefigure}{S\arabic{figure}}
\renewcommand{\thetable}{S\arabic{table}}
\renewcommand{\bibnumfmt}[1]{[S#1]}
\renewcommand{\citenumfont}[1]{S#1}

\begin{figure}[htbp]
\centering
\includegraphics[width=0.4\textwidth]{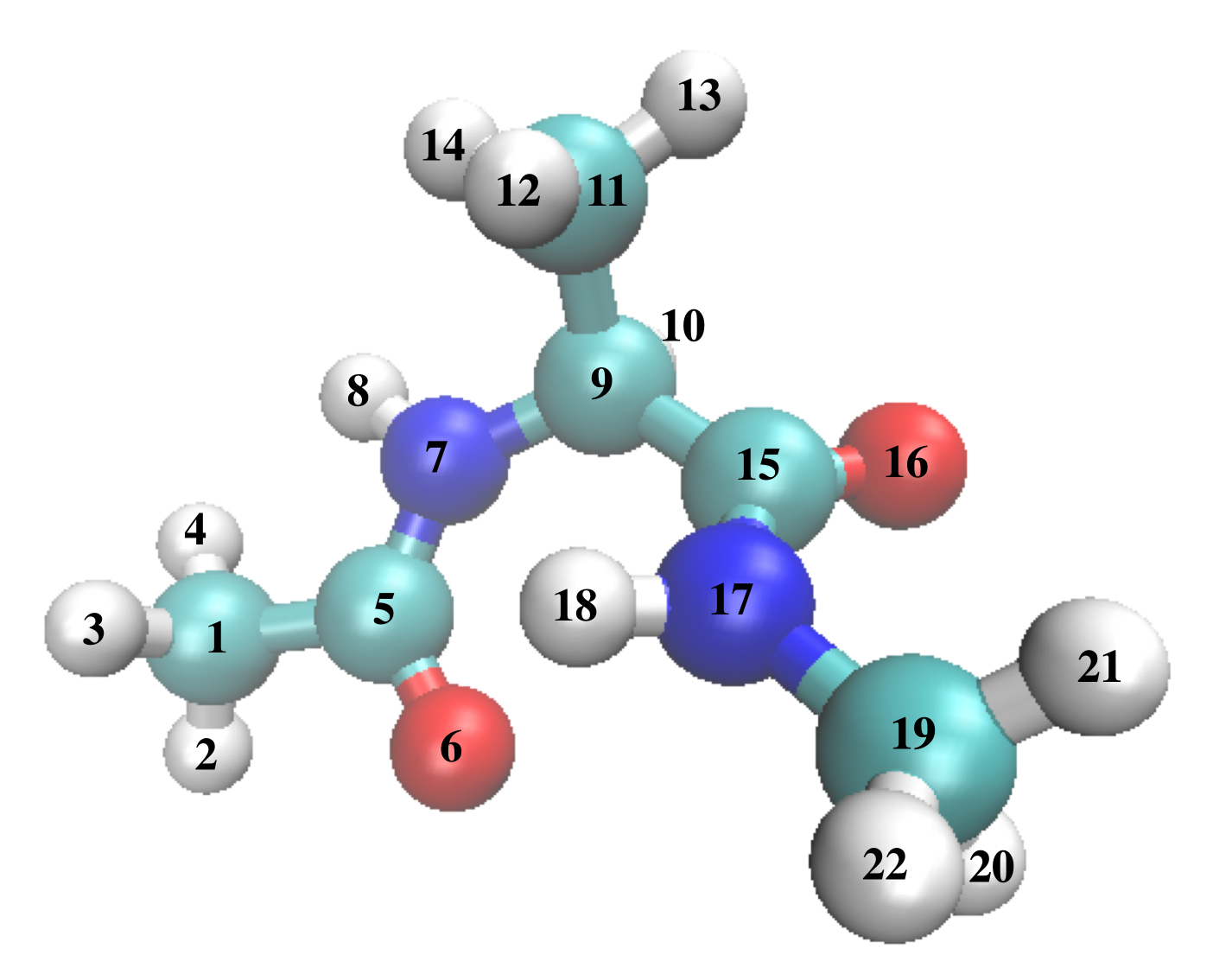}
\caption{Schematic representation of the alanine dipeptide molecule and the numbering of atoms.}
\label{fig:si_mol}
\end{figure}

\begin{table}[htbp]
\caption{Definition of the dihedral angle coordinates corresponding to
 the coefficients $\alpha_i$. The atom numbers are defined in
 Fig. \ref{fig:si_mol}. Note that the dihedral angles are used in cosine
 and sine forms, \textit{i.e.}, $\alpha_{01}$ to $\alpha_{45}$ and
 $\alpha_{46}$ to $\alpha_{90}$ are the cosine and sine forms,
 respectively.}
\centering
\begin{ruledtabular}
      \begin{tabular}{c|lll}
      index & \multicolumn{3}{c}{atom number} \\ \hline
      $01-03$ &  2 -  1 -  5 -  6 &  2 -  1 -  5 -  7  &  3 -  1 -  5 -  6 \\
      $04-06$ &  3 -  1 -  5 -  7 &  4 -  1 -  5 -  6  &  4 -  1 -  5 -  7 \\
      $07-09$ &  1 -  5 -  7 -  8 &  1 -  5 -  7 -  9  &  6 -  5 -  7 -  8 \\
      $10-12$ &  6 -  5 -  7 -  9 &  5 -  7 -  9 - 10  &  5 -  7 -  9 - 11 \\
      $13-15$ &  5 -  7 -  9 - 15 &  8 -  7 -  9 - 10  &  8 -  7 -  9 - 11 \\
      $16-18$ &  8 -  7 -  9 - 15 &  7 -  9 - 11 - 12  &  7 -  9 - 11 - 13 \\
      $19-21$ &  7 -  9 - 11 - 14 & 10 -  9 - 11 - 12  & 10 -  9 - 11 - 13 \\
      $22-24$ & 10 -  9 - 11 - 14 & 15 -  9 - 11 - 12  & 15 -  9 - 11 - 13 \\
      $25-27$ & 15 -  9 - 11 - 14 &  7 -  9 - 15 - 16  &  7 -  9 - 15 - 17 \\
      $28-30$ & 10 -  9 - 15 - 16 & 10 -  9 - 15 - 17  & 11 -  9 - 15 - 16 \\
      $31-33$ & 11 -  9 - 15 - 17 &  9 - 15 - 17 - 18  &  9 - 15 - 17 - 19 \\
      $34-36$ & 16 - 15 - 17 - 18 & 16 - 15 - 17 - 19  & 15 - 17 - 19 - 20 \\
      $37-39$ & 15 - 17 - 19 - 21 & 15 - 17 - 19 - 22  & 18 - 17 - 19 - 20 \\
      $40-42$ & 18 - 17 - 19 - 21 & 18 - 17 - 19 - 22  &  1 -  7 -  5 -  6 \\
      $43-45$ &  5 -  9 -  7 -  8 &  9 - 17 - 15 - 16  & 15 - 19 - 17 - 18 \\
      \end{tabular}
\end{ruledtabular}
\end{table}

\begin{table}[htbp]
\caption{Full list of optimized coefficients for $\lambda=0.5$ in
 descending order. 
The coefficients are calculated as a mean over 10
 optimization trials with different initial parameters, and the standard
 deviations are also calculated from that data.}
\centering
\begin{ruledtabular}
      \begin{tabular}{crc|crc}
      index  & \multicolumn{1}{c}{$\alpha_i$} & \multicolumn{1}{c}{standard deviation} & index  & \multicolumn{1}{c}{$\alpha_i$} & \multicolumn{1}{c}{standard deviation}\\ \hline
      $58$ & $ 1.7453$ & $2.2132 \times 10^{-3}$ & $44$ & $ 0.0893$ & $3.7040 \times 10^{-3}$ \\
      $55$ & $ 1.3872$ & $1.4342 \times 10^{-3}$ & $10$ & $-0.0852$ & $5.4605 \times 10^{-4}$ \\
      $57$ & $-1.2905$ & $1.3520 \times 10^{-3}$ & $78$ & $ 0.0831$ & $1.8153 \times 10^{-3}$ \\
      $12$ & $ 1.1562$ & $1.2566 \times 10^{-3}$ & $90$ & $ 0.0822$ & $4.9167 \times 10^{-5}$ \\
      $11$ & $-1.0431$ & $1.4347 \times 10^{-3}$ & $76$ & $-0.0797$ & $1.5140 \times 10^{-3}$ \\
      $30$ & $-0.9451$ & $1.2216 \times 10^{-3}$ & $62$ & $-0.0794$ & $1.2742 \times 10^{-3}$ \\
      $53$ & $-0.9275$ & $1.2669 \times 10^{-3}$ & $86$ & $ 0.0783$ & $4.0652 \times 10^{-4}$ \\
      $31$ & $ 0.8127$ & $1.2908 \times 10^{-3}$ & $21$ & $ 0.0752$ & $1.6528 \times 10^{-3}$ \\
      $56$ & $-0.4889$ & $1.1470 \times 10^{-3}$ & $40$ & $ 0.0751$ & $6.5192 \times 10^{-4}$ \\
      $33$ & $-0.4320$ & $1.4202 \times 10^{-3}$ & $22$ & $-0.0693$ & $1.4945 \times 10^{-3}$ \\
      $23$ & $-0.4079$ & $1.4942 \times 10^{-3}$ & $03$ & $ 0.0633$ & $1.2441 \times 10^{-3}$ \\
      $70$ & $ 0.3702$ & $1.8008 \times 10^{-3}$ & $37$ & $ 0.0629$ & $3.7637 \times 10^{-4}$ \\
      $87$ & $-0.3522$ & $1.1409 \times 10^{-3}$ & $29$ & $-0.0610$ & $1.1698 \times 10^{-3}$ \\
      $64$ & $ 0.3417$ & $2.4532 \times 10^{-3}$ & $82$ & $-0.0600$ & $7.0894 \times 10^{-4}$ \\
      $59$ & $-0.3412$ & $5.7088 \times 10^{-4}$ & $67$ & $ 0.0529$ & $1.4246 \times 10^{-3}$ \\
      $69$ & $-0.3336$ & $1.4729 \times 10^{-3}$ & $34$ & $ 0.0525$ & $9.6742 \times 10^{-4}$ \\
      $14$ & $-0.3058$ & $4.7081 \times 10^{-4}$ & $66$ & $ 0.0484$ & $1.6093 \times 10^{-3}$ \\
      $74$ & $-0.2931$ & $1.4450 \times 10^{-3}$ & $01$ & $-0.0474$ & $1.2315 \times 10^{-3}$ \\
      $43$ & $-0.2916$ & $3.6195 \times 10^{-4}$ & $50$ & $ 0.0474$ & $1.3886 \times 10^{-3}$ \\
      $17$ & $-0.2855$ & $2.0916 \times 10^{-3}$ & $75$ & $-0.0458$ & $1.9937 \times 10^{-3}$ \\
      $52$ & $-0.2774$ & $1.8192 \times 10^{-3}$ & $06$ & $ 0.0448$ & $1.1690 \times 10^{-3}$ \\
      $25$ & $ 0.2683$ & $1.7937 \times 10^{-3}$ & $20$ & $-0.0446$ & $1.0052 \times 10^{-3}$ \\
      $54$ & $-0.2524$ & $1.8326 \times 10^{-3}$ & $42$ & $-0.0419$ & $2.8231 \times 10^{-3}$ \\
      $60$ & $-0.2518$ & $1.3573 \times 10^{-3}$ & $39$ & $-0.0419$ & $3.9380 \times 10^{-4}$ \\
      $24$ & $ 0.2390$ & $1.5487 \times 10^{-3}$ & $81$ & $ 0.0413$ & $8.6112 \times 10^{-4}$ \\
      $72$ & $ 0.2374$ & $2.9760 \times 10^{-3}$ & $05$ & $-0.0380$ & $1.1829 \times 10^{-3}$ \\
      $88$ & $-0.2369$ & $1.5089 \times 10^{-3}$ & $04$ & $-0.0343$ & $1.2638 \times 10^{-3}$ \\
      $15$ & $-0.2283$ & $1.1347 \times 10^{-3}$ & $89$ & $-0.0342$ & $1.5882 \times 10^{-3}$ \\
      $26$ & $-0.1978$ & $2.2601 \times 10^{-3}$ & $02$ & $-0.0336$ & $1.2283 \times 10^{-3}$ \\
      $08$ & $ 0.1955$ & $1.1138 \times 10^{-3}$ & $45$ & $-0.0329$ & $1.5147 \times 10^{-4}$ \\
      $18$ & $ 0.1869$ & $1.4237 \times 10^{-3}$ & $85$ & $ 0.0328$ & $4.5792 \times 10^{-4}$ \\
      $27$ & $ 0.1803$ & $2.5249 \times 10^{-3}$ & $38$ & $ 0.0318$ & $6.1766 \times 10^{-4}$ \\
      $73$ & $ 0.1789$ & $1.5840 \times 10^{-3}$ & $35$ & $ 0.0302$ & $2.1288 \times 10^{-3}$ \\
      $63$ & $-0.1638$ & $1.4197 \times 10^{-3}$ & $49$ & $-0.0281$ & $1.2168 \times 10^{-3}$ \\
      $65$ & $-0.1600$ & $1.5497 \times 10^{-3}$ & $71$ & $ 0.0266$ & $1.6182 \times 10^{-3}$ \\
      $00$ & $-0.1501$ & $1.7866 \times 10^{-2}$ & $48$ & $-0.0265$ & $1.2725 \times 10^{-3}$ \\
      $28$ & $ 0.1406$ & $1.8230 \times 10^{-3}$ & $19$ & $-0.0254$ & $1.2048 \times 10^{-3}$ \\
      $51$ & $ 0.1394$ & $1.4095 \times 10^{-3}$ & $79$ & $ 0.0222$ & $1.7288 \times 10^{-3}$ \\
      $61$ & $-0.1337$ & $1.5507 \times 10^{-3}$ & $77$ & $-0.0218$ & $1.6835 \times 10^{-3}$ \\
      $16$ & $ 0.1215$ & $3.9594 \times 10^{-4}$ & $80$ & $-0.0165$ & $1.8729 \times 10^{-3}$ \\
      $83$ & $ 0.1181$ & $3.5624 \times 10^{-4}$ & $68$ & $ 0.0147$ & $1.3306 \times 10^{-3}$ \\
      $36$ & $ 0.1160$ & $6.7586 \times 10^{-4}$ & $32$ & $ 0.0083$ & $1.1014 \times 10^{-3}$ \\
      $46$ & $ 0.1109$ & $8.6875 \times 10^{-4}$ & $09$ & $ 0.0060$ & $2.2018 \times 10^{-4}$ \\
      $47$ & $-0.1089$ & $8.5633 \times 10^{-4}$ & $41$ & $-0.0049$ & $5.2831 \times 10^{-4}$ \\
      $13$ & $-0.1017$ & $1.3712 \times 10^{-3}$ & $07$ & $-0.0049$ & $2.2382 \times 10^{-4}$ \\
      $84$ & $ 0.1005$ & $1.0053 \times 10^{-3}$ &               &                          &                         \\ 
      \end{tabular}
\end{ruledtabular}
\end{table}

\begin{table}[htbp]
\caption{Full list of optimized coefficients for $\lambda=0$ in
 descending order.
The coefficients are calculated as a mean over 10 optimization
 trials with different initial parameters, and the standard deviations
 are also calculated from that data.}
\centering
\begin{ruledtabular}
      \begin{tabular}{crc|crc}
      index & \multicolumn{1}{c}{$\alpha_i$} & \multicolumn{1}{c}{standard deviation} & index  & \multicolumn{1}{c}{$\alpha_i$} & \multicolumn{1}{c}{standard deviation}\\ \hline
      $58$ & $ 3.1511$ & $5.1786 \times 10^{-3}$ & $61$ & $ 0.4045$ & $2.5976 \times 10^{-3}$ \\
      $55$ & $ 2.6830$ & $1.4032 \times 10^{-2}$ & $78$ & $-0.4023$ & $3.4432 \times 10^{-2}$ \\
      $33$ & $-2.5446$ & $4.1285 \times 10^{-3}$ & $83$ & $ 0.3947$ & $4.6369 \times 10^{-4}$ \\
      $57$ & $-2.4873$ & $2.1762 \times 10^{-3}$ & $01$ & $ 0.3868$ & $7.7135 \times 10^{-3}$ \\
      $00$ & $-2.4527$ & $1.1267 \times 10^{-2}$ & $63$ & $-0.3619$ & $3.0409 \times 10^{-2}$ \\
      $31$ & $ 2.1781$ & $2.3398 \times 10^{-2}$ & $19$ & $-0.3565$ & $1.4991 \times 10^{-2}$ \\
      $72$ & $ 1.9539$ & $2.0216 \times 10^{-2}$ & $21$ & $-0.3389$ & $3.2438 \times 10^{-2}$ \\
      $88$ & $-1.9151$ & $6.4681 \times 10^{-4}$ & $73$ & $-0.3162$ & $3.2698 \times 10^{-2}$ \\
      $30$ & $-1.5587$ & $2.4409 \times 10^{-2}$ & $02$ & $ 0.3098$ & $7.6417 \times 10^{-3}$ \\
      $51$ & $ 1.4762$ & $1.0109 \times 10^{-2}$ & $22$ & $ 0.3030$ & $2.7601 \times 10^{-2}$ \\
      $50$ & $ 1.3540$ & $1.0037 \times 10^{-2}$ & $66$ & $ 0.2748$ & $1.8121 \times 10^{-2}$ \\
      $11$ & $-1.2952$ & $5.0231 \times 10^{-3}$ & $56$ & $-0.2683$ & $2.3946 \times 10^{-3}$ \\
      $87$ & $ 1.2813$ & $1.2720 \times 10^{-2}$ & $25$ & $ 0.2247$ & $1.5989 \times 10^{-2}$ \\
      $23$ & $-1.2131$ & $2.6910 \times 10^{-2}$ & $17$ & $-0.2183$ & $3.6273 \times 10^{-2}$ \\
      $74$ & $-1.1767$ & $3.2587 \times 10^{-2}$ & $38$ & $-0.2111$ & $6.6435 \times 10^{-4}$ \\
      $64$ & $ 1.1428$ & $2.3731 \times 10^{-2}$ & $85$ & $ 0.1958$ & $4.1448 \times 10^{-4}$ \\
      $60$ & $-1.1346$ & $5.5465 \times 10^{-4}$ & $40$ & $ 0.1867$ & $6.7666 \times 10^{-4}$ \\
      $13$ & $-1.0363$ & $2.4769 \times 10^{-3}$ & $53$ & $ 0.1794$ & $1.3984 \times 10^{-2}$ \\
      $35$ & $-1.0152$ & $5.7957 \times 10^{-3}$ & $29$ & $-0.1559$ & $6.2233 \times 10^{-3}$ \\
      $26$ & $-1.0069$ & $3.1832 \times 10^{-2}$ & $39$ & $ 0.1538$ & $4.4290 \times 10^{-4}$ \\
      $67$ & $ 0.9278$ & $1.7525 \times 10^{-2}$ & $42$ & $-0.1359$ & $7.2953 \times 10^{-3}$ \\
      $54$ & $-0.9158$ & $2.3805 \times 10^{-3}$ & $20$ & $ 0.1330$ & $9.2173 \times 10^{-4}$ \\
      $52$ & $-0.9051$ & $2.3491 \times 10^{-3}$ & $41$ & $-0.1272$ & $6.2702 \times 10^{-4}$ \\
      $71$ & $ 0.8552$ & $1.9949 \times 10^{-2}$ & $82$ & $ 0.1268$ & $6.4682 \times 10^{-4}$ \\
      $12$ & $ 0.8358$ & $2.3220 \times 10^{-3}$ & $28$ & $-0.1038$ & $6.6050 \times 10^{-3}$ \\
      $46$ & $ 0.7259$ & $9.4158 \times 10^{-3}$ & $07$ & $ 0.1030$ & $6.6437 \times 10^{-4}$ \\
      $65$ & $-0.6920$ & $2.6471 \times 10^{-2}$ & $75$ & $-0.0929$ & $1.7938 \times 10^{-2}$ \\
      $70$ & $ 0.6651$ & $3.4239 \times 10^{-2}$ & $69$ & $ 0.0900$ & $2.3117 \times 10^{-2}$ \\
      $15$ & $ 0.6425$ & $1.1172 \times 10^{-3}$ & $08$ & $-0.0854$ & $1.9408 \times 10^{-3}$ \\
      $47$ & $ 0.5937$ & $9.4431 \times 10^{-3}$ & $36$ & $ 0.0847$ & $3.3590 \times 10^{-4}$ \\
      $80$ & $-0.5795$ & $3.4149 \times 10^{-2}$ & $90$ & $ 0.0836$ & $4.5581 \times 10^{-5}$ \\
      $49$ & $ 0.5640$ & $6.9340 \times 10^{-3}$ & $43$ & $-0.0797$ & $2.6093 \times 10^{-4}$ \\
      $84$ & $ 0.5562$ & $7.6356 \times 10^{-4}$ & $14$ & $-0.0766$ & $1.3332 \times 10^{-3}$ \\
      $10$ & $-0.5522$ & $2.8041 \times 10^{-3}$ & $68$ & $-0.0761$ & $1.4963 \times 10^{-2}$ \\
      $76$ & $ 0.5451$ & $1.8858 \times 10^{-2}$ & $32$ & $-0.0745$ & $2.0748 \times 10^{-3}$ \\
      $62$ & $ 0.5373$ & $1.3239 \times 10^{-2}$ & $09$ & $ 0.0496$ & $6.3746 \times 10^{-4}$ \\
      $48$ & $ 0.5277$ & $7.1636 \times 10^{-3}$ & $37$ & $ 0.0462$ & $3.7478 \times 10^{-4}$ \\
      $81$ & $ 0.5218$ & $6.9539 \times 10^{-4}$ & $77$ & $-0.0438$ & $3.0750 \times 10^{-3}$ \\
      $24$ & $ 0.5206$ & $2.0542 \times 10^{-2}$ & $79$ & $ 0.0288$ & $3.1391 \times 10^{-3}$ \\
      $27$ & $ 0.5132$ & $3.3043 \times 10^{-2}$ & $34$ & $-0.0206$ & $1.9660 \times 10^{-3}$ \\
      $04$ & $-0.4836$ & $1.1262 \times 10^{-2}$ & $05$ & $ 0.0176$ & $9.6553 \times 10^{-3}$ \\
      $44$ & $ 0.4811$ & $3.2205 \times 10^{-3}$ & $89$ & $ 0.0165$ & $3.1549 \times 10^{-2}$ \\
      $18$ & $ 0.4788$ & $1.1043 \times 10^{-2}$ & $16$ & $-0.0136$ & $6.3249 \times 10^{-4}$ \\
      $59$ & $-0.4441$ & $1.2642 \times 10^{-3}$ & $06$ & $ 0.0112$ & $9.4690 \times 10^{-3}$ \\
      $86$ & $ 0.4393$ & $6.8539 \times 10^{-4}$ & $45$ & $-0.0090$ & $1.3192 \times 10^{-4}$ \\
      $03$ & $-0.4289$ & $1.1060 \times 10^{-2}$ &               &                          &                         \\ 
      \end{tabular}
\end{ruledtabular}
\end{table}

\begin{table}[htbp]
\caption{Full list of optimized coefficients for $\lambda=10$ in descending
 order. The coefficients are calculated as a mean over 10 optimization
 trials with different initial parameters, and the standard deviations
 are also calculated from that data.}
\centering
\begin{ruledtabular}
      \begin{tabular}{crc|crc}
      index & \multicolumn{1}{c}{$\alpha_i$} & \multicolumn{1}{c}{standard deviation} & index & \multicolumn{1}{c}{$\alpha_i$} & \multicolumn{1}{c}{standard deviation}\\ \hline
      $58$ & $ 0.8306$ & $3.5528 \times 10^{-11}$ & $81$ & $-0.0426$  & $1.2914 \times 10^{-10}$ \\
      $57$ & $-0.6144$ & $2.7921 \times 10^{-9} $ & $01$ & $-0.0422$  & $6.8381 \times 10^{-11}$ \\
      $12$ & $ 0.6061$ & $2.3452 \times 10^{-9} $ & $04$ & $-0.0379$  & $3.6555 \times 10^{-10}$ \\
      $55$ & $ 0.5980$ & $6.3127 \times 10^{-10}$ & $45$ & $-0.0372$  & $2.2530 \times 10^{-9} $ \\
      $11$ & $-0.5796$ & $1.5295 \times 10^{-9} $ & $18$ & $ 0.0372$  & $3.6965 \times 10^{-9} $ \\
      $53$ & $-0.3841$ & $3.6147 \times 10^{-10}$ & $68$ & $ 0.0352$  & $3.5254 \times 10^{-9} $ \\
      $88$ & $ 0.3787$ & $1.0132 \times 10^{-10}$ & $65$ & $-0.0350$  & $1.5159 \times 10^{-10}$ \\
      $30$ & $-0.2475$ & $1.5217 \times 10^{-9} $ & $50$ & $ 0.0349$  & $1.7245 \times 10^{-10}$ \\
      $56$ & $-0.2391$ & $3.0540 \times 10^{-9} $ & $09$ & $ 0.0317$  & $1.2569 \times 10^{-9} $ \\
      $61$ & $-0.2257$ & $3.9281 \times 10^{-10}$ & $51$ & $ 0.0316$  & $1.4641 \times 10^{-10}$ \\
      $87$ & $-0.2200$ & $3.5622 \times 10^{-10}$ & $20$ & $ 0.0313$  & $3.2898 \times 10^{-9} $ \\
      $15$ & $-0.2189$ & $3.5264 \times 10^{-10}$ & $84$ & $ 0.0293$  & $1.8027 \times 10^{-10}$ \\
      $31$ & $ 0.2153$ & $1.7952 \times 10^{-9} $ & $34$ & $ 0.0292$  & $1.8532 \times 10^{-9} $ \\
      $00$ & $ 0.2084$ & $1.0023 \times 10^{-7} $ & $64$ & $ 0.0280$  & $3.1097 \times 10^{-9} $ \\
      $26$ & $ 0.1699$ & $2.0812 \times 10^{-9} $ & $41$ & $-0.0280$  & $6.1336 \times 10^{-11}$ \\
      $27$ & $-0.1531$ & $1.8384 \times 10^{-9} $ & $83$ & $ 0.0267$  & $2.9348 \times 10^{-10}$ \\
      $52$ & $-0.1359$ & $8.2253 \times 10^{-11}$ & $02$ & $-0.0258$  & $7.9951 \times 10^{-11}$ \\
      $43$ & $-0.1239$ & $9.3277 \times 10^{-10}$ & $80$ & $-0.0249$  & $9.6400 \times 10^{-10}$ \\
      $76$ & $-0.1054$ & $3.8599 \times 10^{-9} $ & $66$ & $ 0.0240$  & $2.8176 \times 10^{-9} $ \\
      $75$ & $ 0.0983$ & $3.9213 \times 10^{-9} $ & $40$ & $ 0.0238$  & $1.2308 \times 10^{-10}$ \\
      $13$ & $ 0.0930$ & $3.2933 \times 10^{-9} $ & $05$ & $-0.0232$  & $2.7483 \times 10^{-10}$ \\
      $71$ & $ 0.0898$ & $3.4497 \times 10^{-9} $ & $89$ & $-0.0231$  & $3.0971 \times 10^{-10}$ \\
      $60$ & $ 0.0847$ & $5.6629 \times 10^{-10}$ & $33$ & $-0.0227$  & $4.5047 \times 10^{-9} $ \\
      $37$ & $ 0.0800$ & $2.0658 \times 10^{-10}$ & $10$ & $-0.0213$  & $3.0889 \times 10^{-9} $ \\
      $25$ & $ 0.0775$ & $3.7010 \times 10^{-9} $ & $17$ & $-0.0197$  & $1.4371 \times 10^{-9} $ \\
      $54$ & $-0.0764$ & $1.8622 \times 10^{-10}$ & $22$ & $-0.0177$  & $1.9466 \times 10^{-9} $ \\
      $70$ & $ 0.0752$ & $8.2912 \times 10^{-10}$ & $85$ & $-0.0147$  & $2.2215 \times 10^{-10}$ \\
      $69$ & $-0.0711$ & $2.7979 \times 10^{-9} $ & $74$ & $ 0.0146$  & $4.9195 \times 10^{-11}$ \\
      $19$ & $-0.0684$ & $2.2126 \times 10^{-9} $ & $24$ & $ 0.0146$  & $2.4496 \times 10^{-9} $ \\
      $23$ & $-0.0677$ & $1.2557 \times 10^{-9} $ & $08$ & $-0.0137$  & $3.4721 \times 10^{-9} $ \\
      $03$ & $ 0.0676$ & $2.7891 \times 10^{-10}$ & $63$ & $ 0.0136$  & $4.5148 \times 10^{-10}$ \\
      $39$ & $-0.0646$ & $1.6649 \times 10^{-10}$ & $21$ & $ 0.0129$  & $1.4208 \times 10^{-9} $ \\
      $07$ & $-0.0615$ & $1.2193 \times 10^{-9} $ & $79$ & $-0.0117$  & $7.3789 \times 10^{-10}$ \\
      $49$ & $-0.0595$ & $2.4875 \times 10^{-10}$ & $29$ & $ 0.0086$  & $4.0031 \times 10^{-9} $ \\
      $72$ & $-0.0594$ & $3.5546 \times 10^{-9} $ & $32$ & $ 0.0083$  & $1.9369 \times 10^{-9} $ \\
      $06$ & $ 0.0588$ & $3.5111 \times 10^{-11}$ & $48$ & $-0.0078$  & $2.5259 \times 10^{-11}$ \\
      $82$ & $-0.0554$ & $5.5851 \times 10^{-10}$ & $14$ & $ 0.0066$  & $3.6773 \times 10^{-10}$ \\
      $78$ & $ 0.0527$ & $6.9169 \times 10^{-10}$ & $42$ & $-0.0064$  & $4.6971 \times 10^{-9} $ \\
      $90$ & $ 0.0504$ & $8.0539 \times 10^{-10}$ & $77$ & $ 0.0059$  & $1.0523 \times 10^{-9} $ \\
      $86$ & $ 0.0495$ & $3.2697 \times 10^{-11}$ & $36$ & $ 0.0057$  & $2.3601 \times 10^{-10}$ \\
      $62$ & $-0.0472$ & $3.4940 \times 10^{-9} $ & $35$ & $ 0.0035$  & $4.2877 \times 10^{-9} $ \\
      $16$ & $ 0.0455$ & $8.4477 \times 10^{-10}$ & $28$ & $-0.0035$  & $3.9852 \times 10^{-9} $ \\
      $46$ & $ 0.0442$ & $1.4667 \times 10^{-11}$ & $59$ & $-0.0026$  & $7.3267 \times 10^{-10}$ \\
      $47$ & $-0.0432$ & $1.8039 \times 10^{-10}$ & $38$ & $ 0.0022$  & $4.3125 \times 10^{-10}$ \\
      $73$ & $-0.0431$ & $3.1798 \times 10^{-10}$ & $44$ & $-0.0017$  & $4.6810 \times 10^{-9} $ \\
      $67$ & $-0.0430$ & $2.7864 \times 10^{-9} $ &               &                          &                         \\
      \end{tabular}
\end{ruledtabular}
\end{table}

\begin{figure}[htbp]
\centering
\includegraphics[width=0.8\textwidth]{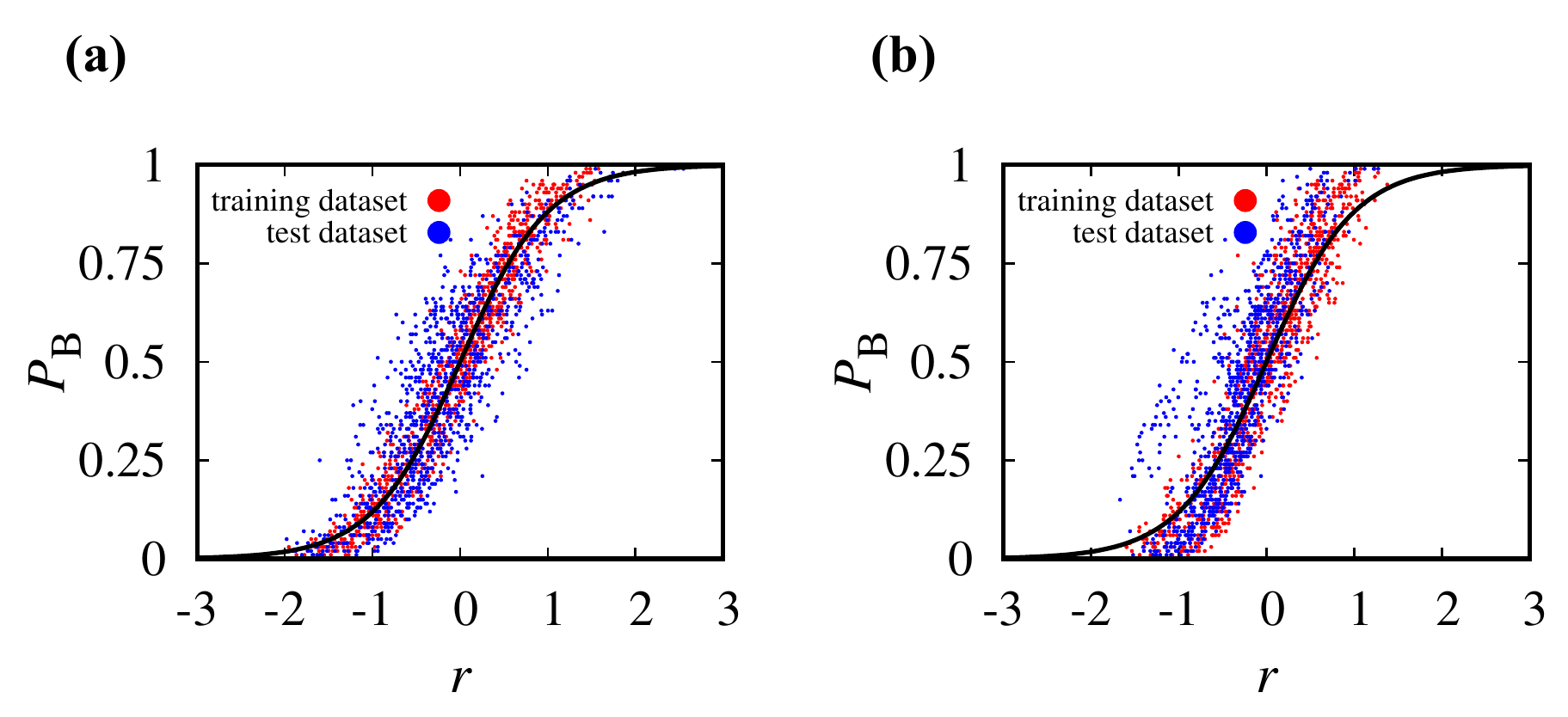}
\caption{Committor distributions of the training (red) and test (blue)
 datasets as functions of the optimized coordinate $r$
 for the cases of (a) $\lambda=0$ and (b) $\lambda=10$. 
The sigmoid function $p_\mathrm{B}(r)=[1+\tanh(r)]/2$ is shown in black line.}
\end{figure}

\begin{figure}[htbp]
\centering
\includegraphics[width=0.8\textwidth]{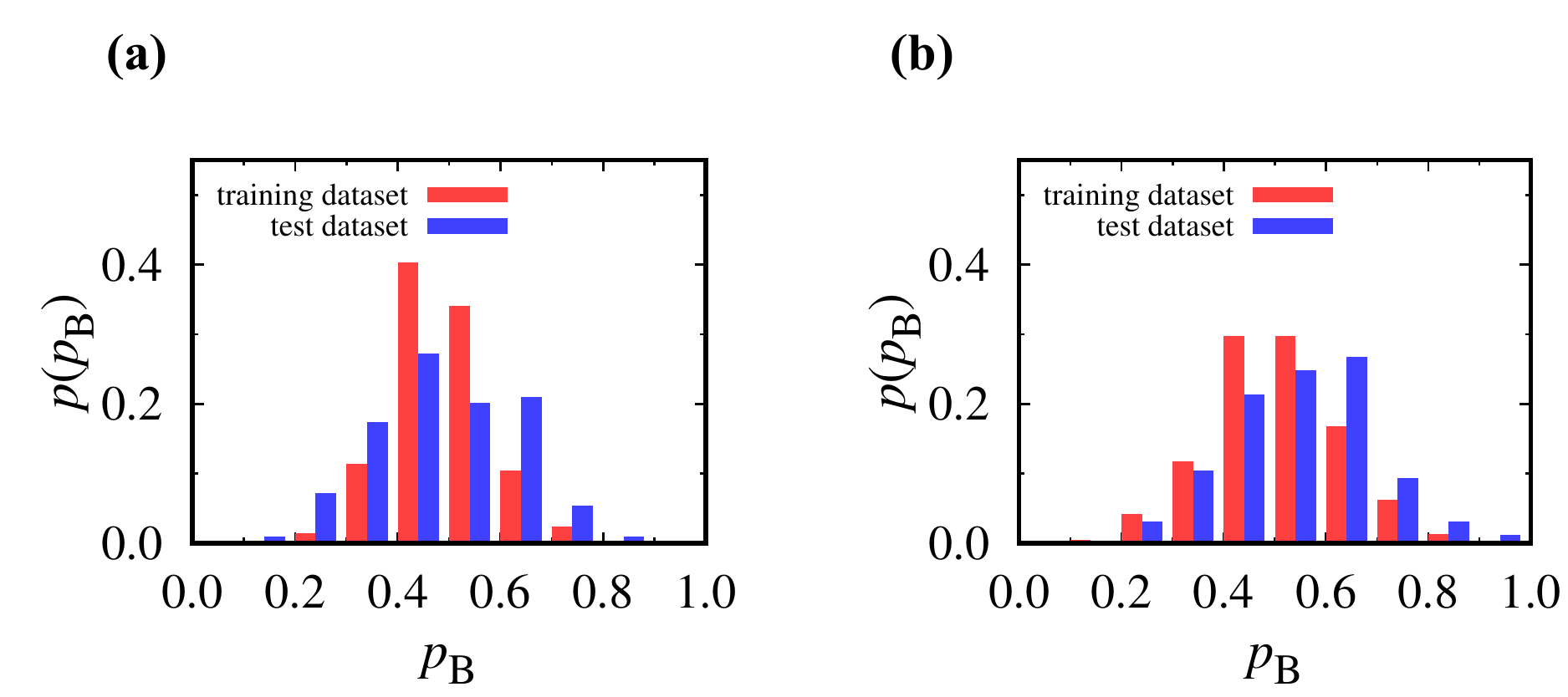}
\caption{Probability of $p_\mathrm{B}$ at about the transition state of $r$
 ($-0.2 \le r \le 0.2$) for the cases of (a) $\lambda=0$ and (b)
 $\lambda=10$. Red and blue bars are for training and testing datasets,
 respectively.}
\end{figure}

\end{document}